\documentclass[superscriptaddress,aps,letterpaper,10pt,twocolumn,floats,showpacs,amsmath,amsfonts,amssymb,pre,nofootinbib]{revtex4-2}

\usepackage[utf8]{inputenc}

\usepackage{url,psfrag,graphicx}
\usepackage{dcolumn}
\usepackage{amsmath,amssymb}
\usepackage{bm}
\usepackage{hyperref}
\usepackage{float,epsfig,color}
\usepackage{times}

\usepackage{array}
\usepackage{braket}
\usepackage{diagbox}
\usepackage{multirow}
\usepackage{stackrel}

\usepackage{verbatim}
\usepackage{xcolor}
\usepackage{soul}

\DeclareMathAlphabet      {\mathitbf}{OML}{cmm}{b}{it}

\begin{document}

\title{Microscopic fluctuations in the spreading fronts of circular wetting liquid droplets}

\author{J. M. Marcos}
\affiliation{Departamento de F\'{\i}sica, Universidad de Extremadura, 06006 Badajoz, Spain}
\address{Instituto de Computaci\'on Cient\'{\i}fica Avanzada de Extremadura (ICCAEx), Universidad de Extremadura, 06006 Badajoz, Spain}
\author{J. J. Mel\'endez}
\affiliation{Departamento de F\'{\i}sica, Universidad de Extremadura, 06006 Badajoz, Spain}
\address{Instituto de Computaci\'on Cient\'{\i}fica Avanzada de Extremadura (ICCAEx), Universidad de Extremadura, 06006 Badajoz, Spain}
\author{R. Cuerno}
\affiliation{Departamento de Matem\'aticas and Grupo Interdisciplinar de Sistemas Complejos (GISC), Universidad Carlos III de Madrid, 28911 Legan\'es, Spain}
\author{J. J. Ruiz-Lorenzo}
\affiliation{Departamento de F\'{\i}sica, Universidad de Extremadura, 06006 Badajoz, Spain}
\address{Instituto de Computaci\'on Cient\'{\i}fica Avanzada de Extremadura (ICCAEx), Universidad de Extremadura, 06006 Badajoz, Spain}

\date{\today}

\begin{abstract}
We study numerically the kinetic roughening properties of the precursor fronts of nonvolatile liquid droplets spreading on solid substrates, for the case of circular droplets, more frequently addressed in experiments. To this end, we perform kinetic Monte Carlo (kMC) simulations of a lattice gas model whose kinetic roughening behavior has been recently assessed in a band geometry [J.\ M.\ Marcos {\em et al.}, Phys.\ Rev.\ E {\bf 105}, 054801 (2022)]. We compare the scaling behaviors of the spreading fronts obtained for the two geometries, in view of the occurrence of, for example, different universality subclasses for different growth geometries for the related  important Kardar-Parisi-Zhang (KPZ) universality class. For circular droplets we obtain that the average front position increases (sub-)diffusively as $R\sim t^{\delta}$, where $\delta \lesssim 1/2$ shows a stronger dependence on the conditions considered for temperature and substrate wettability than in band geometry. In spite of this, front fluctuations for circular droplets behave qualitatively similar to those seen for band geometries, with kinetic roughening exponent values which similarly depend on temperature $T$ but become $T$-independent for sufficiently high $T$. Circular droplets also display intrinsic anomalous scaling with different values of the roughness exponent at short and large length scales, and fluctuations statistics which are close to the Tracy-Widom probability distribution function that applies in the corresponding KPZ universality subclass, now the one expected for interfaces with an overall circular symmetry.
\end{abstract}

\maketitle

\section{Introduction}\label{sec:intro}

Under complete wetting conditions, non-volatile fluids spreading on top of solid substrates \cite{Starov2019,Reiter2018} are known to develop so-called precursor fronts at the edge of ultrathin (molecular-depth) fluid layers \cite{Popescu2012}, that preempt the spreading of the macroscopic droplets. Given such small length scales, the dynamics of these precursor fronts is strongly influenced by fluctuations. Their average position $R(t)$ displays diffusive-like behavior with time $t$ as $R(t)\sim t^{\delta}$ with values of $\delta$ definitely larger (i.e., faster spreading) than those characterizing the radius of the macroscopic droplet, namely, Tanner's [e.g., $\delta_{\rm Tanner}=1/10$ in three dimensions (3D)] \cite{Tanner1979} and related growth laws \cite{Bonn2009}. Hence, the propagation of precursor fronts needs to be taken into account when probing fluid motion at nanoscales, as recently assessed e.g.\ in the imbibition of liquids in nanoporous photonic crystals \cite{Cencha2020} or in graphene nanochannels \cite{Chang2024}.

The fluctuations that become so relevant at the small scales characteristic of precursor films do indeed enhance their dynamics, as is also the case, for example, for the free surface of very thin viscous fluid layers \cite{Nesic2015,Zhao2022}; but they also impact their correlations in space, inducing the scale-invariant behavior characteristic of kinetically rough surfaces and interfaces \cite{Barabasi1995,Krug1997,Abraham2002,Zhang2021}. Thus, these fronts evolve in absence of typical space-time scales analogous to, for instance, an equilibrium system at the critical temperature of a continuous transition \cite{Tauber2014}, with the difference being that for many kinetically rough surfaces there is no need to tune system parameters to precise critical values \cite{Barabasi1995,Krug1997}. In general, kinetic roughening implies power-law behavior of system observables in time and space, characterized by critical exponents with values that can be classified into universality classes, akin to those of equilibrium critical phenomena. Actually, surface kinetic roughening has unraveled scaling behavior which generalizes the latter. For instance, the dynamic scaling ansatz satisfied by the classical models A or B of critical dynamics \cite{Tauber2014} (termed Family-Viseck (FV) in the kinetic roughening context \cite{Barabasi1995,Krug1997}) generalizes into so-called anomalous scaling \cite{Sarma1994,Plischke1993,Schroeder1993,Lopez1997,Ramasco2000}, found in many models and experiments of rough interfaces \cite{Cuerno2004,Cuerno2007}. Also, and related with recent developments largely connected with the paramount Kardar-Parisi-Zhang (KPZ) universality class \cite{Kriecherbauer2010,Takeuchi2018}, additional traits are proving to bear significance on the unambiguous identification of the universality class. Specifically, the probability distribution function (PDF) of rescaled front fluctuations around the mean, which happens to take on a universal form. For 1D interfaces in the KPZ class, and sufficiently far in time both, from the initial conditions and from saturation to steady state, it is given by some member of the Tracy-Widom (TW) family of PDFs \cite{Fortin2015}, the precise one depending on boundary conditions. Thus, for interfaces in a band geometry and periodic boundary conditions (PBC) it is the TW PDF associated with the largest eigenvalue of random matrices in the Gaussian orthogonal ensemble (TW-GOE), which is replaced by that related with the Gaussian unitary ensemble (TW-GUE) for interfaces with an overall circular geometry \cite{Kriecherbauer2010,Takeuchi2018}. The latter has been recently found to account for the fluctuations of an astounding variety of rough interfaces in terms of physical nature and typical scales \cite{Makey2020}.

Back to fronts of spreading droplets, it is actually the circular geometry which is most frequently observed in experiments and atomistic [e.g., molecular dynamics (MD)] models \cite{Bonn2009,Popescu2012}. However, the fluctuation (kinetic roughening) properties of spreading fronts have been addressed only in the band geometry thus far. Specifically, by resorting to kinetic Monte Carlo (kMC) simulations ---with improved access over MD models to the large space-time scales required to unambiguously identify scaling behavior---, a lattice model has been studied in Ref.\ \cite{Marcos2022} with Hamiltonian \cite{Abraham2002,Marcos2022}
\begin{equation}
    \mathcal{H}= -J \sum_{\langle \mathitbf{r}, \mathitbf{s} \rangle} n(\mathitbf{r},t)n(\mathitbf{s},t) - A \sum_{\mathitbf{r}}\frac{n(\mathitbf{r},t)}{Z^3},
    \label{eq:energy}
\end{equation}
where $n(\mathitbf{r},t)$ is the occupation number of lattice sites by the fluid. The first term describes the interactions between the liquid particles and their nearest neighbors, characterized by the coupling constant $J$ that quantifies the strength of this interaction and it is related to the surface tension between the fluid and vacuum, while the second one accounts for the interaction between the fluid and the substrate, characterized by a Hamaker constant $A > 0$, and $\mathitbf{r}=(x,y,Z)$ denotes position in a three-dimensional (3D) cubic lattice. Note that the present is a statistical (rather than fully atomistic) description of the fluid in which a particle is to be understood, rather, as a group of fluid molecules rather than a single one, with presence/absence of a particle at a lattice site representing an enhanced/decreased probability of local occurrence of fluid molecules there, see a discussion e.g.\ in Ref.\ \cite{Chalmers2017} and other therein. The advantage of this modeling approach is that, even if simplified, it enables the efficient study of the interfacial scaling properties of the system while correctly including the relevant structure and thermodynamics of the fluid \cite{Areshi2019}.

The model defined through Eq.\ \eqref{eq:energy} was first introduced to describe spreading of liquid droplets by Lukkarinen {\em et al.} \cite{Lukkarinen1995}, and the fluctuation properties of the fronts of the precursor and supernatant films that ensue (at $Z=1$ and 2, respectively) have since been studied in the band geometry \cite{Abraham2002,Harel2018,Harel2021,Marcos2022}, where PBC are applied along one of the two substrate directions and the macroscopic droplet is a reservoir of particles placed on a segment along the perpendicular direction. The main conclusions are \cite{Marcos2022} that: ({\em i}) kinetic roughening properties of both, supernatant and precursor films are the same, with $\delta\approx 0.5$ basically for all $(A,T)$ parameter choices; ({\em ii}) the fronts display intrinsics anomalous scaling with critical exponent values 
which depend more strongly on $T$ than on $A$, becoming $T$-independent for sufficiently high temperatures; and ({\em iii}) the fluctuation PDF is the TW-GOE, as for the 1D KPZ equation on a band geometry with PBC. This last result and the approximate global exponent values agreed with similar results for a continuum model put forward for high $T$ conditions \cite{Abraham2002,Marcos2022} which, notably, displays a KPZ nonlinearity. However, the coupling of such a term is proportional to the front speed $V(t) \sim t^{\delta-1}$, hence it does not control the values of the critical exponents at long times while (quite remarkably) simultaneously inducing TW fluctuation statistics.

Although the original model proposed by Lukkarinen {\em et al}.\ \cite{Lukkarinen1995} had multiple layers along the vertical direction, they showed that the main mechanism for the growth of the precursor film are: ({\em i}) holes in the precursor film that travel backwards towards the macroscopic droplet and ({\em ii}) particles located in the supernatant layer that diffuse until they reach the edge of the precursor film or some hole of the precursor layer to fill. For that reason, later models that study the precursor film \cite{Abraham2002} restrict themselves to two layers only.

In this paper we study numerically the kinetic roughening properties of the fronts of fluid circular droplets spreading on solid substrates, as described by the lattice gas with Hamiltonian given by Eq.\ \eqref{eq:energy}. At this, our goal is to provide data which replicate more closely typical experimental setups in this context \cite{Bonn2009,Popescu2012} and potentially present analogous complexities and challenges to ours in terms of data analysis. Notably, to the best of our knowledge, there are very few, if any, reports in the literature on intrinsic anomalous scaling for interfaces with a circular geometry. Additionally, through comparison with the results of Ref.\ \cite{Marcos2022}, here we address the dependence of scaling behavior with system geometry that has been naturally introduced recently in kinetic roughening by the KPZ universality class \cite{Kriecherbauer2010,Takeuchi2018}, which happens to be directly relevant to our system in band geometry.

This paper is organized as follows. After this introduction, Sec.\ \ref{observables} contains a description of our numerical simulation method for the lattice gas model defined by Eq.\ \eqref{eq:energy}, including the definition of the various observables measured. Our numerical results are then reported, together with a discussion, in Sec.\ \ref{sec:results}. A summary of our results, together with our conclusions, appears in Sec.\ \ref{sec:concl}. Finally, further additional simulation details, tables, and results are provided in Appendix \ref{details}.

\section{Simulation details and definitions} \label{observables}
\subsection{MC Simulation details}
The discrete driven Ising lattice gas model considered herein comprises two overlapping $2D$ rectangular layers, of dimensions $L_{x} \times L_{y}$. Without loosing generality, we choose $L_{x} = L_{y}\equiv L_{\rm side}$ throughout herein, 
with $L_{\rm side}$ being an odd number. Each node of the cubic lattice $\mathitbf{r}= (x, y, Z)$ can be occupied by at most one particle at any time; consequently, the occupation number $n(\mathitbf{r},t)$ may take on the values 0 for an empty site or 1 for an occupied one. The lower $(Z = 1)$ and upper $(Z=2)$ layers are referred to as the precursor and supernatant layers, respectively. The substrate on top of which the droplet expands is located at $Z = 0$.

\begin{figure}[t]
\centering
\includegraphics[width=0.45\textwidth]{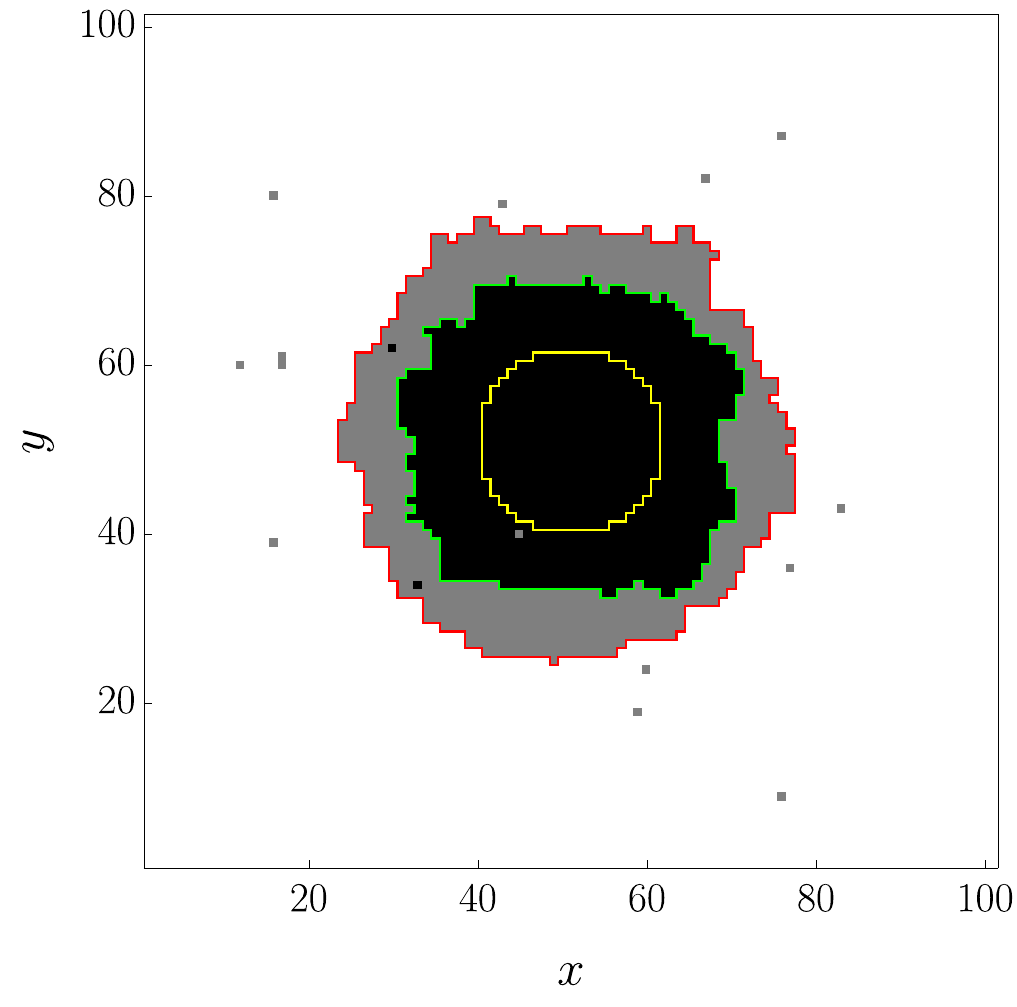}
\caption{(Color online) Top view of a snapshot of the lattice gas model. The cells that are occupied in the precursor and supernatant layers are represented in gray and black, respectively, while empty cells are uncolored. The red and green lines delimit the corresponding fronts, while the yellow line delimits the reservoir. The conditions used were $T=1/3$, $A=10$, $R_R=11$, $J=1$, and $L_{\rm side}=101$.}
\label{fig:Snapshot}
\end{figure}

In order to simulate the dynamics of the system, it is necessary to define the fluid reservoir, which represents the macroscopic droplet that feeds the growth of the films. 
Initially, only these cells are occupied; if due to an exchange (the dynamics we have used is described in detail below) any of the cells of the reservoir becomes empty, it is instantaneously refilled. In a previous study on spreading in a band or rectangular geometry \cite{Marcos2022}, the reservoir was the first $(x=0)$ column of both layers. In the present case, it can be chosen in several ways. The most straightforward approach is to define the reservoir as the central cell of the system. This choice, however, results in a markedly slow growth of the fluid films, as there is only a single cell in each layer capable of feeding the film growth and the development of a film takes prohibitively long simulation times. On the other hand, a point-like reservoir seems too strong of an idealization regarding experimental conditions. An alternative approach was taken, whereby a larger reservoir is employed comprising all the cells located at a distance smaller than a specified radius $R_{R}$ (radius of the reservoir) from the center of the system. The shape of such a reservoir is illustrated in Figure \ref{fig:Snapshot}. This choice results in a markedly faster system dynamics. Further discussion on the size and shape of the reservoir can be found in Appendix~\ref{reservoir}. Conversely, if a particle occupies the outer boundary of the lattice at a later stage, the particle is assumed to escape from the system.

The boundary condition we employ is closest to the situation for e.g.\ spontaneous imbibition of a fluid into a porous medium, which corresponds to having a fluid reservoir at constant pressure. This condition is well known to lead to Washburn’s diffusive law for the time increase of the column of invading fluid (and time decreasing average velocity of the invading front), analogous in our case to the $R(t)\sim t^{\delta}$ diffusive law for the time increase of the extent of the precursor film. In contrast, a fixed-flux boundary condition (as implied e.g.\ by injecting fluid with a syringe) leads to a time-independent average velocity for the invading front; see e.g.\ Ref.\ \cite{Alava2004}.

The total energy of the system is given by Eq.\ \eqref{eq:energy}, defined in terms of $A, J >0$. From the physical point of view \cite{Bonn2009,Popescu2012}, the most interesting values for the pairs $(A,J)$ are those for which $J/k_{\mathrm{B}}T$ is sufficiently large to achieve a high degree of involatility and $A/k_{\mathrm{B}}T$ is sufficiently large to be in the complete wetting regime, as outlined in Ref.\ \cite{Abraham2002}. Therefore, of all conditions reported in this work (see Table \ref{tab:parametros}), the most realistic and therefore the closest to those of liquids that present a precursor film are those in which $A$ is large and $T$ is low. In this work, we also report results for conditions that do not meet the above criteria since we are interested in studying the model by itself over a wide range of parameters, comparing the results with those obtained in the band geometry, where the system was also studied over a similarly wide range of parameters. From Eq.\ \eqref{eq:energy}, the lowest energy is clearly attained for the smallest value of $Z$, which suggests that full occupation of the precursor layer is energetically favorable. This preferential occupation is enhanced for $A \gg J$, in which case one would expect the bottom layer to grow at a faster rate than the upper one. On the contrary, when $J$ is dominant, both layers are be expected to grow at similar speeds. Henceforth, we will choose physical units such that $k_{\mathrm{B}}=1$ and remain arbitrary otherwise.

The evolution of the system has been simulated by continuous-time Monte Carlo Kawasaki local dynamics \cite{Newman1999}; for further details, the reader is referred to Appendix A of Ref.\ \cite{Marcos2022}. At each time, a particle is classified as belonging to the precursor (or the supernatant) film if there are nearest-neighbor connections filled with particles all the way back to the droplet reservoir. Moreover, a particle is considered to be at the front if it belongs to the film and there exists an empty nearest-neighbor cell connected to the system boundary through empty nearest-neighbors. It is evident that this ``strict'' definition of the front is computationally very inefficient. In order to enhance the efficiency of the algorithm, an alternative, simplified definition of the front has been used herein. In particular, we do not verify whether there are connections extending to the system boundary, which is situated at a distance much greater than the typical film sizes in our simulations. Instead, we only check if there are empty nearest-neighbors connections up to a distance that is twice the last size measurement for the film plus an offset of 10. We assume that if there is a path up to this distance, there will also be a path to the system boundary. In Fig.\ \ref{fig:Snapshot}, the ensuing fronts of the precursor film and the supernatant film appear in red and green lines, respectively. 

Table \ref{tab:parametros} shows a summary of all the simulations performed, showing the parameters of each simulation and the number of realizations used in each one.

\begin{table}[!htbp]
\begin{ruledtabular}
\begin{tabular}{p{1cm}|p{1cm}|p{1cm}|p{2.5cm}|p{2cm}}
$L_{\rm side}$ & $T$ & $A$ & $N_E$ & Number of runs \\ \hline\hline
  & 10 & 10 & $1.25 \times 10^8$ & 50 \\ 
    & 10 & 3 & $1.25 \times 10^8$ & 50 \\ 
1001  & 10 & 1 & $2.5 \times 10^8$ & 50 \\ 
    & 10 & 1/3 & $2.5 \times 10^8$ & 50 \\ 
      & 10 & 0.1 & $2.5 \times 10^8$ & 50 \\ \hline

   & 3 & 10 & $2.5 \times 10^8$ & 50 \\ 
    & 3 & 3 & $2.5 \times 10^8$ & 100 \\ 
1001  & 3 & 1 & $5 \times 10^8$ & 100 \\ 
    & 3 & 1/3 & $5 \times 10^8$ & 100 \\ 
      & 3 & 0.1 & $5 \times 10^8$ & 100 \\ \hline

   & 1 & 10 & $2.5 \times 10^8$ & 100 \\ 
    & 1 & 3 & $2.5 \times 10^8$ & 100 \\ 
1001  & 1 & 1 & $5 \times 10^8$ & 112 \\ 
    & 1 & 1/3 & $5 \times 10^8$ & 125 \\ 
      & 1 & 0.1 & $5 \times 10^8$ & 125 \\ \hline

   & 3/4 & 10 & $5 \times 10^8$ & 125 \\ 
    & 3/4 & 3 & $5 \times 10^8$ & 125 \\ 
1001 & 3/4 & 1 & $2.5 \times 10^9$ & 150 \\ 
    & 3/4 & 1/3 & $2.5 \times 10^9$ & 150 \\ 
      & 3/4 & 0.1 & $2.5 \times 10^9$ & 150 \\ \hline

   & 1/2 & 10 & $1.25 \times 10^9$ & 125 \\ 
    & 1/2 & 3 & $2.5 \times 10^9$ & 150 \\ 
1001  & 1/2 & 1 & $5 \times 10^9$ & 150 \\ 
    & 1/2 & 1/3 & $5 \times 10^9$ & 150 \\ 
      & 1/2 & 0.1 & $5 \times 10^9$ & 150 \\ \hline

   & 1/3 & 10 & $2.5 \times 10^9$ & 100 \\ 
    & 1/3 & 3 & $5 \times 10^9$ & 150 \\ 
1001  & 1/3 & 1 & $7.5 \times 10^9$ & 100 \\ 
    & 1/3 & 1/3 & $7.5 \times 10^9$ & 100 \\ 
      & 1/3 & 0.1 & $7.5 \times 10^9$ & 100 \\ 
\end{tabular}
\end{ruledtabular}
\caption{Parameters used for the runs reported herein. Here, $N_E$ represents the total number of the exchanges performed, and the last column shows the number of runs launched in each case.}
\label{tab:parametros}
\end{table}

\subsection{Front position and roughness}

The average distance of the precursor or supernatant interface from the center of the system at a time $t$ is defined as
\begin{equation}
    \bar{h}(t,Z)=\frac{1}{N_Z}\sum_{i}h_{i}(t,Z)\,,
    \label{eq:distanciaProm1}
\end{equation}
where $N_Z$ is the number of points that belong to the corresponding front, $h_{i}(t,Z)$ are the Euclidean distances from the cells of that front to the center of the reservoir and the sum spans all cells $i$ belonging to the front as defined above. It should be noted that $N_Z$ differs for the two layers and changes over time. Likewise, the average distance from the interface to the fluid reservoir (the normalized average front position) is given by \cite{Huergo2012}
\begin{equation}
    \overline{h_R}(t,Z)=\frac{1}{N_Z}\sum_{i}\left(h_{i}(t,Z)-R_{R}\right)\,.
    \label{eq:distanciaProm2}
\end{equation}

On the other hand, the front width (or roughness) at each layer, $w(N_Z,t,Z)$, is defined as the standard deviation of the corresponding front values \cite{Barabasi1995,Krug1997}, namely,
\begin{equation}
	\label{eq:width}
	w^2(N_Z,t,Z)=\left\langle \overline{[h_{i}(t,Z)-\bar{h}(t,Z)]^2} \right\rangle,
\end{equation}
where we denote $\overline{O}(t,Z) \equiv (1/N_Z)\sum_i O_i(t,Z)$ for the average of a given observable $O_i(t,Z)$ defined at the position of the front on each layer. Furthermore, we denote by $\langle (\cdots) \rangle$ the average over different realizations (noise) of the system. Hereafter, we will denote this observable as $w^2(N_Z,t)$, or simply as $w^2(t)$. Also, we will omit the layer $Z$ index, whose value will be clear from the context.

\begin{figure}[t]
\centering
\includegraphics[width=0.45\textwidth]{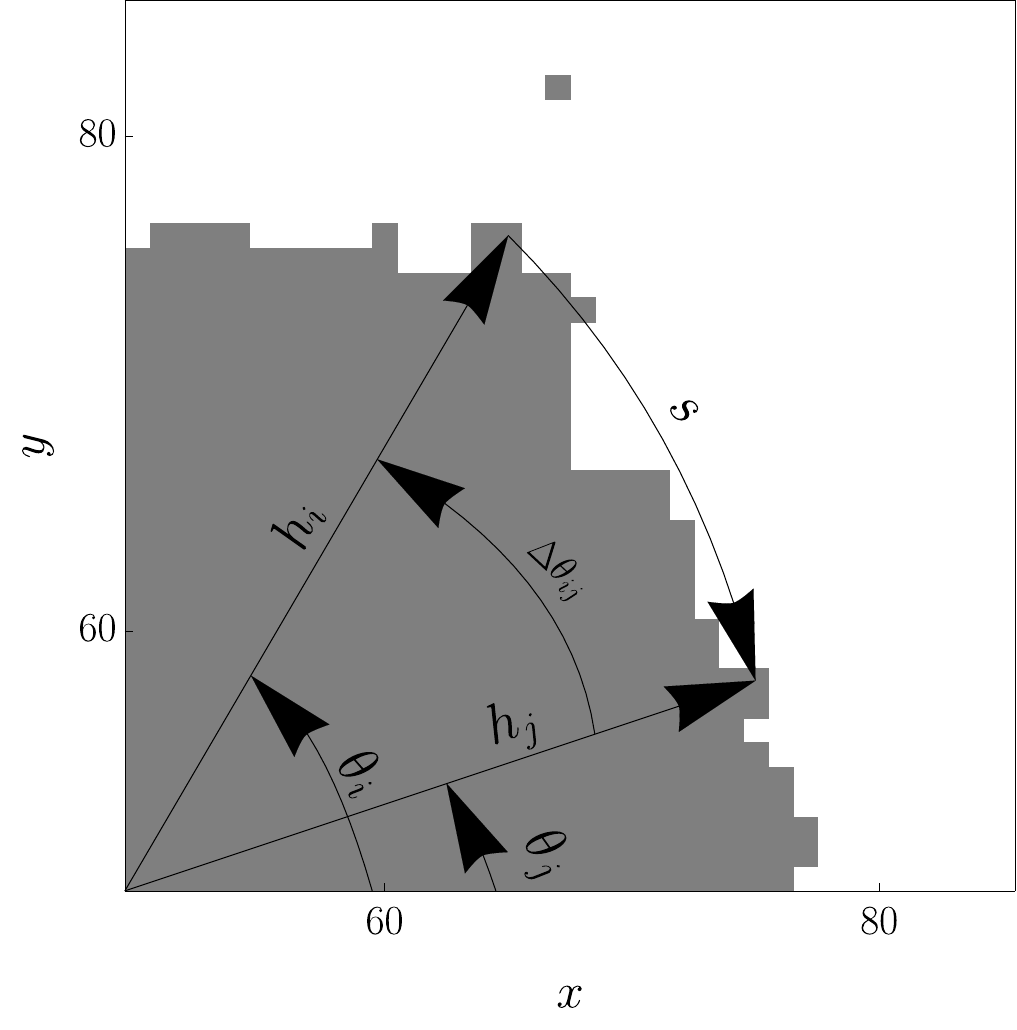}
\caption{Zoom of Fig.\ \ref{fig:Snapshot} showing the distances to the center of the reservoir $h_i$ and $h_j$, the angles $\theta_i$ and $\theta_j$, the angle difference $\Delta\theta_{ij}$ between two cells $i$ and $j$ belonging to the front of the precursor, and the arc length $s = \bar h \Delta\theta_{ij}$.}
\label{fig:GrafIRS}
\end{figure}

\subsection{Height-difference correlation function}

In order to describe the dynamical evolution of the front it is pertinent to study the height-difference correlation function, defined as: 
\begin{equation}
	\label{eq:correlation_2}
    	C_2(s, t)=\frac 1{N} \sum_{\bar{h}\Delta\theta_{ij}\in s} \left\langle [h_{i}(t)-h_j(t)]^2 \right\rangle,
\end{equation}
where $\Delta\theta_{ij}=\left(\theta_i-\theta_j\right) \text{mod } 2\pi$ is the angular difference between the cells $i$ and $j$ and $s \equiv \bar{h}\Delta\theta_{ij}$ is therefore the arc length between these cells, see Fig.~\ref{fig:GrafIRS}. Thus, the sum spans all the pairs of cells whose arc length is $s$. In Eq.\ \eqref{eq:correlation_2}, $N$ is the number of those pairs. By definition, $s$ takes values between $0$ and $2 \pi \bar{h}(t)=L_f(t)$, where $L_f(t)$ is the average front length. It should be noted that $L_f(t)$ grows with time, as $\bar{h}(t)$ also increases. As there are many possible arc differences between cells, we compute the value of the function $C_2(s,t)$ by discretizing the angle interval $[0,2\pi)$ (and thus the arc length interval) in boxes $(\theta-\delta\theta,\theta+\delta\theta)$ where $\delta\theta$ is a parameter that sets the width of the interval. In practice, we set $\delta\theta$ as $\delta\theta=2\pi/N_{\mathrm{A,B}}$, where $N_{\mathrm{A,B}}$ sets the number of angular boxes (bins) in which we discretize the interval $[0,2\pi)$. The particular choice for $N_{\mathrm{A,B}}$ does not change the results obtained.
This analysis has already been used to study the radial growth of experimental cell colonies \cite{Galeano2003,Huergo2011,Huergo2012,Santalla2018} and tumors \cite{Bru1998,Bru2003,Block2007}, and for both continuous \cite{Santalla2014,Santalla2015,Santalla2018} and discrete \cite{Santalla2018b} models of surface kinetic roughening.

The height-difference correlation function, Eq.\ \eqref{eq:correlation_2}, quantifies the statistical correlations between the front positions at points separated a given arc length $s$ along the interface. The term 'height' does not refer to the physical height of the film but originates from deposition models, such as ballistic deposition, which were among the first to study kinetic roughening of interfaces \cite{Baiod1988,Meakin1986,Barabasi1995}. In our case, we use 'height' to describe the position of the front, which is a function of both arc length and time. Since $C_2(s,t)$ is defined as the squared difference between height values at locations separated by a distance $s$, it is generally expected to increase with $s$ (the farther away $h_i$ and $h_j$ are, the larger their difference will be) until reaching a saturation value.


As noted in Sec.\ \ref{sec:intro}, under the present conditions for fluid film spreading the average front (i.e., our estimate for the radius $R(t)$ of the precursor layer for $Z=1$) is expected to grow as $ \overline{h_R}(t)\sim t^\delta$, with $\delta = 1/2$ for all parameter conditions \cite{Harel2018,Abraham2002,Marcos2022}. On the other hand, previous works on our lattice gas model in the band geometry \cite{Abraham2002,Harel2018,Marcos2022} also show that kinetic roughening conditions hold, whereby the roughness $w(L_f,t)$ satisfies the FV dynamic scaling ansatz \cite{Barabasi1995,Krug1997,Huergo2012}, i.e.,
\begin{equation}
	\label{eq:w}
	w(L_f,t)=t^{\beta}f\left( t/L_f^z \right),
\end{equation}
according to which the roughness increases as $w\sim t^{\beta}$ for relatively short times such that $t \ll L_f^z$, while it saturates to $w_{\mathrm{sat}}\sim L_f^{\alpha}$ for long enough times such that $t \gg L_f^z$. Here, $\alpha=\beta z$ is the global roughness exponent, which is related to the fractal dimension of the front \cite{Barabasi1995,Mozo2022} and characterizes the fluctuations of its roughness at (large) scales comparable with the system size.
Further, $\beta$ in Eq.\ \eqref{eq:w} denotes the growth exponent and $z$ is the so-called dynamic exponent, which quantifies the power-law increase of the lateral correlation length $\xi(t)$ along the front \cite{Barabasi1995,Krug1997},
\begin{equation}
    \label{eq:correlation_length}
    \xi(t) \sim t^{1/z}\,.
\end{equation}
For our system, the previous discussion should be refined. Note that, in this context in which the front length $L_f$ is growing, two correlation lengths can be defined. A correlation length $\xi_\infty(t)$, defined in a system in which the front length $L_f$ is infinite, and $\xi_L(t)$, defined in a finite-sized front. The first always grows as $\xi_\infty(t)\sim t^{1/z}$ while the latter grows as $\xi_L(t)\sim t^{1/z}$ except when it reaches the length of the front $L_f$. In this case $\xi_L(t)\sim L_f$, i.e. it grows at the same pace as the front length.

With this clarification we can redefine the roughness as:
\begin{equation}
    \label{eq:w_x2}
    w(L_f,t)=t^{\beta}f\left( \xi_\infty(t)/L_f (t)\right),
\end{equation}
where $f(u)\sim \mathrm{const}$ for $u\ll1$, i.e., increases as $w\sim t^{\beta}$ for times such that $\xi_\infty (t)\ll L_f(t)$ and $f(u)\sim u^{-\alpha}$ for $u\gg 1$, i.e. the system roughness only saturates to $L_f^\alpha$ for those times in which the $\xi_\infty (t)\gg L_f(t)$, i.e. those times in which the finite correlation length $\xi_L(t)$ has reached the front length $L_f$ and grows with it. In our simulations we did not find any evidence of saturation in which the $\xi_L(t)$ reached the front length. Therefore, from here onwards, we will use simply $\xi(t)$ when referring to the correlation length. Kinetic roughening is indeed a generalization of equilibrium critical dynamics \cite{Tauber2014}, where space-time scale criticality emerges as a result of the power-law increase of the correlation length with time. Saturation to a steady state, as for the front roughness, takes place once $\xi(t)$ reaches the system size $L$. However, for growing fronts with a circular geometry, $L_f(t)$ frequently grows approximately linearly with time \cite{Huergo2011,Huergo2012,Bru1998,Bru2003} while $z>1$, so that saturation does not occur as $\xi(t) < L$ for all times. For the case presented here, $L_f$ grows in time at the same rate as $\bar{h}$. We have found no evidence of saturation to a steady-state value in our simulations.

\subsection{Scaling behavior of the height-difference correlation function}

In the context of scaling behavior, the function $C_2(s,t)$, which is quadratic in the height, is expected to scale with the lateral distance as
\begin{equation}
    \label{eq:correlation_length_loc}
    C_2(s,t) \sim s^{2\alpha_{\rm loc}}\,,
\end{equation}
for distances smaller than the correlation length, i.e.\ such that $s\ll\xi(t)$. In this context, $\alpha_{\rm loc}$ refers to a roughness exponent measured at local distances, smaller than the system size. Under the FV scaling hypothesis, $\alpha_{\rm loc}=\alpha$ \cite{Barabasi1995,Krug1997}. However, there are more complex scaling scenarios, termed anomalous scaling \cite{Sarma1994,Plischke1993,Schroeder1993,Krug1997,Lopez1997,Ramasco2000,Cuerno2004,Marcos2022}, for which $\alpha_{\rm loc}\neq\alpha$. As will be discussed below, the results of our kMC simulations are consistent in particular with so-called intrinsic anomalous scaling, for which the height-difference correlation function behaves as \cite{Lopez1997}
\begin{equation}
    C_2(s,t)=s^{2\alpha} g(s/\xi(t)) \,,
    \label{eq:corr_length}
\end{equation}
where 
$g(u) \sim u^{-2\alpha^\prime}$ for $u\ll 1$ and $g(u) \sim u^{-2\alpha}$ for $u\gg 1$, with $\alpha^\prime \equiv \alpha-\alpha_\mathrm{loc}$. 
Standard FV scaling corresponds to $\alpha'=0$, so that $g_{\rm FV}(u) \sim$ constant for $u\ll1$ \cite{Barabasi1995,Krug1997}. In this FV case, the surface is a self-affine fractal and there is a single roughness exponent which characterizes both small- and large-scale fluctuations in space. In contrast, under intrinsic anomalous scaling, the condition of strict self-affinity is not fulfilled, and the local and global space fluctuations do not scale with the same exponent. It should be noted that in this case there are three independent exponents (rather than two, $\alpha$ and $z$, for FV scaling) characterizing the scaling behavior, e.g.\, $\alpha$, $\alpha_{\rm loc}$, and $z$ \cite{Lopez1997}.

As will be discussed below, the existence of two distinct low- and high-temperature morphological regimes for our fronts 
requires the introduction of two alternative approaches for computing the correlation length. In the simplest cases, once the function $C_2(s,t)$ reaches a plateau as a function of $s$ for a fixed time $t$, 
Eq.\ \eqref{eq:corr_length} allows one to compute the correlation length $\xi_a(t)$ at that time \cite{Barreales2020} through
\begin{equation}
C_2(\xi_a(t),t)= a \, C_2^\mathrm{plateau}(t)\,,
\label{eq:c2_scaling}
\end {equation}
where $a$ is a constant, typically $a \gtrsim 0.8$. In other words, the correlation length at a given time $t$ is defined as the distance along the front at which the correlation function $C_2$ takes on the $a$ fraction of its plateau value $C_2^\mathrm{plateau}(t)$. The specific value of $a$ does not affect the scaling of the correlation length \cite{Barreales2020,Marcos2022}.

For more complex scenarios in which the function $C_2(s,t)$ oscillates, 
we instead approximate the first peak of the function as a parabola through a weighted least squares fit using the points that are above of the $90\%$ of the maximum value of the peak. Subsequently, the plateau value is approximated as the maximum value of the parabola, and the correlation length is defined as the arc length that reaches $90\%$ of the plateau value. Appendix~\ref{parabola} includes further details on this procedure.

\subsection{Morphology of the system and local observables}

\begin{figure*}[t]
\centering
\includegraphics[width=0.96\textwidth]{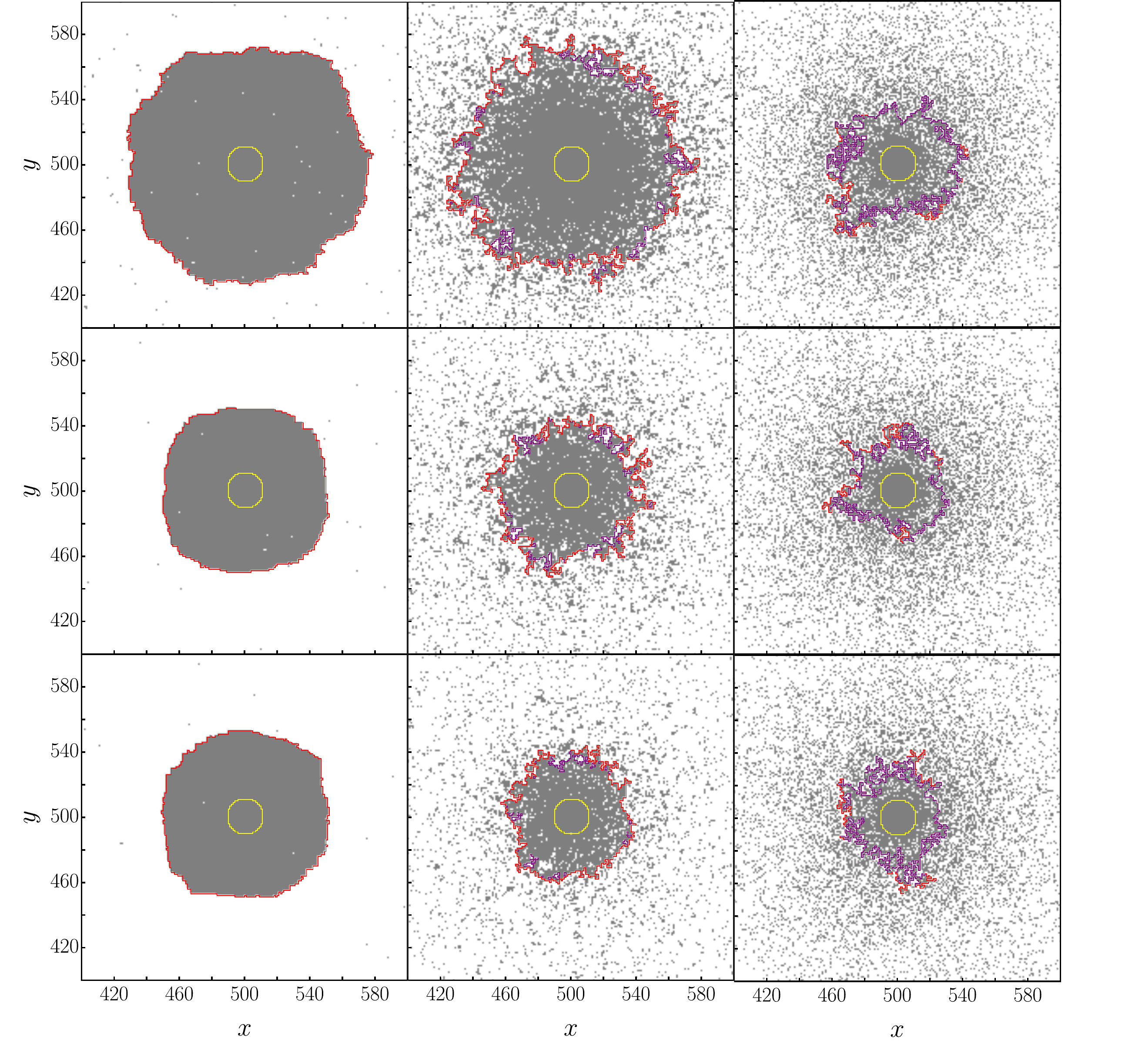}
\caption{(Color online) Top views of snapshots of the bottom layer ($Z=1$) of the lattice gas model. Occupied cells are in gray, while empty cells are uncolored. The red line delimits the front of the precursor film computed in accordance with the strict definition, whereas the purple line delimits the front as computed through the eased alternative definition. The yellow line delimits the reservoir. The conditions used were $T=\{1/3,1,10\}$, $A=\{10,1,0.1\}$, $R_R=11$, $J=1$, and $L_{\rm side}=1001$. $T$ increases from left to right and $A$ increases from bottom to top.}
\label{fig:morfologia}
 \end{figure*}

Figure \ref{fig:morfologia} shows the morphology of the expanding precursor film of the droplet for a few conditions studied. The front computed according to the ``strict'' definition is shown in red, whereas we show in purple the front computed by allowing diagonal neighbors in the process of searching for empty cells connected to the system boundary (i.e., ``eased'' definition of the front). This figure evidences that the behavior of the system undergoes a significant transformation with increasing temperature. For high temperatures, the thermal fluctuations in the system are typically much larger than the cohesive energy of the liquid particles. In this case, the particles tend to diffuse instead of grouping together. Specifically, at very high temperatures, the system exhibits a considerable degree of noise, which challenges the unambiguous definition of a front. In some extreme cases, this may even result into the inability to find any point belonging to the front. For that reason, we have excluded from our study very large values of the temperature, and only results up to $T=3$ will be presented and discussed. The front shape for the conditions closest to the experiments is illustrated in the upper left corner of Fig.\ \ref{fig:morfologia}.

Figure \ref{fig:morfologia} also illustrates how, at intermediate temperatures at which a front can be defined but lacks connectivity and it is more dispersed, the front presents some gaps when computed from its strict definition. In contrast, the eased definition results into a closed front with no gaps, although eventually some additional extraneous points might add up. 

Our simulation results and observables employ the strict
definition of the front. As gaps in the front are more prevalent at
higher temperatures, we have verified explicitly (not shown) that the
exponents obtained from both definitions of the front are the same for
one of such conditions ($T = A = 1$). We are confident that the
reported exponents remain consistent for other parameter values. It is
important to note that the gaps that develop in the front using the
strict definition are relatively few and small and appear at different locations over time, so that their effect is expected to vanish when averaging over multiple runs. Note that the conditions shown in the rightmost column of Fig.\ \ref{fig:morfologia} correspond to $T=10$ and for these conditions we have refrained from giving results since the front could not be defined clearly.


Futhermore, Fig.\ \ref{fig:morfologia} also shows that at very low temperature and low Hamaker constant the shape of the film deviates from the circular form, taking a square-like configuration. The relation between this shape and the peaks eventually exhibited by the height-difference correlation function, as well as the underlying cause of these peaks, will be discussed in detail in subsequent sections. In order to gain a deeper understanding of these specific cases where the shape of the film is not circular, we will define the roughness 
with reference to a local front. This approach has been recently used, for instance, in the study of discrete models of first-passage percolation,
see Ref.\ \cite{Domenech2024} and others therein. 
Namely, we define the average front position in an angular box $\Omega$ as:
\begin{equation}
    \bar{h}_\Omega(t,Z)=\frac{1}{N_Z(\Omega)}\sum_{i\in\Omega}h_{i}(t,Z)\,,
    \label{eq:distanciaPromANG}
\end{equation}
where $\Omega=(\theta-\delta\theta,\theta+\delta\theta)$, $\delta\theta$ is the parameter used to discretize the $[0,2\pi)$ interval and $N_Z(\Omega)$ is the number of points that belong to the corresponding front in the angular box $\Omega$. The front width is defined for these cases as
\begin{equation}
	\label{eq:width_shape}
	w^2_{\Omega}(L_f,t,Z)=\left\langle \overline{[h_{i}(t,Z)-\left\langle\bar{h}_\Omega(t,Z)\right\rangle]^2} \right\rangle,
\end{equation}
where $\bar{h}_\Omega(t,Z)$ is the average \eqref{eq:distanciaPromANG} taken in the angular box into which the cell $i$ falls. This new definition of the front width will result into a different value for the growth exponent, denoted as $\beta_{\Omega}$. The definition of the height-difference correlation function remains unchanged, and therefore only one value is presented for the exponents $\alpha$ and $z$.

The uncertainties of the fluctuations and the correlation functions have been calculated following the jackknife procedure \cite{Young2015,Efron1982}. Since the algorithm we use is a continuous-time Monte Carlo, each run has different times. To perform the averaging of the magnitudes, we define time-boxes in which we perform the averages. Further details about the use of the jackknife procedure and the time-boxes are included in Appendix B of Ref.\ \cite{Barreales2020}. In this work we use time-boxes which are evenly spaced in a logarithmic representation of time.

\section{Results and discussion} \label{sec:results}

In this section we will analyze the evolution of the different observables described in the previous section. For simplicity we will only present the Figures and Tables corresponding to the precursor film. The exponents Tables for the supernatant film can be found in Appendix \ref{extra}.

\subsection{Front position and roughness}\label{sec:RW}

\begin{table*}[t]
\begin{ruledtabular}
\begin{tabular}{|c|c|c|c|c|c|c|c|c|c|c|}
\hline 
\multirow{2}{*}{\diagbox[width=2.5em]{$A$}{$T$}} & \multicolumn{2}{c|}{3}  & \multicolumn{2}{c|}{1}  & \multicolumn{2}{c|}{3/4}  & \multicolumn{2}{c|}{1/2} & \multicolumn{2}{c|}{1/3} \\  \cline{2-11}

 & $\delta$ & $2\beta$& $\delta$ & $2\beta$& $\delta$ & $2\beta$& $\delta$ & $2\beta$& $\delta$ & $2\beta$ \\ \hline
 \hline
    10 & 0.394(2) & 0.463(9)& 0.426(1) & 0.512(8) & 0.4360(8) &0.50(1) &0.4672(8) &0.25(2) & * & 0.20(6)\\ \hline
    3 & 0.368(2) & 0.420(6) & 0.413(1) & 0.50(1) & 0.430(1) & 0.51(1) & 0.446(1) & 0.24(4) & * & 0.17(6)\\ \hline
    1 & 0.336(3) & 0.41(1) &  0.374(2) & 0.48(1) & 0.367(4) & 0.3(1) & 0.3488(9) & 0.24(3) & 0.3493(8) & 0.14(2) \\ \hline
    1/3 & 0.345(2) & 0.400(7) & 0.369(2) & 0.44(1) & 0.378(3) & 0.23(4) & 0.350(1) & 0.20(3) & 0.3441(8) & 0.17(3) \\ \hline
    0.1 & 0.341(3) & 0.398(6) & 0.372(2) & 0.44(1) & 0.382(2) & 0.27(2) & 0.3534(8) & 0.24(4) & 0.3501(9) & 0.14(2)\\ 
 \hline
\end{tabular}
\end{ruledtabular}
\caption{Values of the exponents $\delta$ and $2\beta$ for the precursor layer for all the conditions under study. The two conditions in which the average front position does not reach a regime governed by a power law are indicated with an asterisk.}
\label{tab:delta_precursor}
\end{table*}

\begin{figure}[t]
\centering
\includegraphics[width=0.45\textwidth]{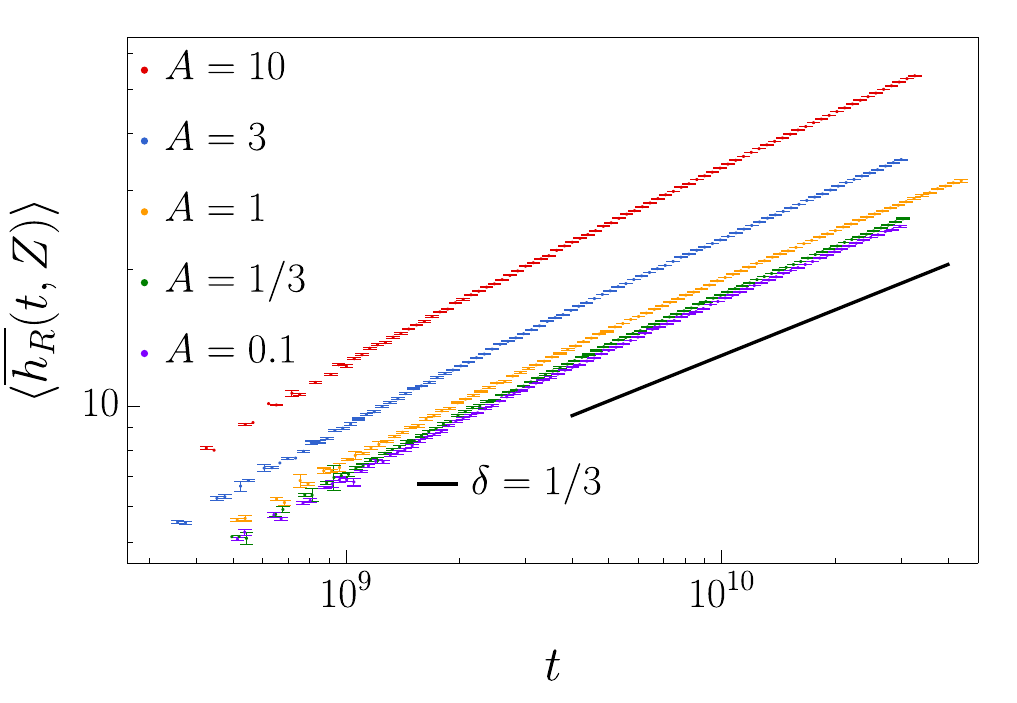}
\caption{(Color online) Average front position $\langle \overline{h_R}(t,Z)\rangle$ plotted as a function of time for $T=3$, $R_R=11$, $Z=1$, and several values of $A$, see legend. The solid black line corresponds to the reference scaling $\langle \overline{h_R}(t,Z)\rangle \sim t^{1/3}$. All units are arbitrary in this and all figures below.}
\label{fig:ht}
\end{figure}

Figure \ref{fig:ht} shows the behavior of $\langle \overline{h_R}(t,Z)\rangle$ for five different values of the Hamaker constant. For virtually all values of $A$ and $T$ the mean front position grows as $\langle \overline{h_R}(t) \rangle \sim t^\delta$ at long times, as expected, with exponent values depending on the parameters, as detailed in Table \ref{tab:delta_precursor} (and also in Table \ref{tab:delta_supernatant}, provided in Appendix~\ref{extra}) for the precursor and supernatant layers, respectively. 
In a previous work \cite{Marcos2022}, in which the same system was simulated in a band geometry, the scaling exponent $\delta$ was found to be approximately equal to $1/2$ for all the conditions studied and for both layers. In contrast, in a circular geometry $\delta$ takes values between $1/3$ and $1/2$ for most conditions. Subdiffusive non-Tanner values for $\delta$ have also been obtained in MD simulations of circular fluid droplets, see, for instance, Ref.\ \cite{Weng2017}. In the cases of $T=1/3$ and $A=10$ and 3 (which represent relevant conditions in terms of precursor spreading, as the ratios $J/k_{\mathrm{B}}T$ and $A/k_{\mathrm{B}}T$ take on their largest values), the average front position $\overline{h_R}(t)$ does not reach a regime governed by a power law (as a reference, see Fig.\ \ref{fig:ht_anomalo} in Appendix \ref{extra} for $A =10$). Therefore, for these two cases we have not reported a value for the $\delta$ exponent. However, it should be noted that the behavior of $\langle \overline{h_R}(t) \rangle$ does not differ much from the anticipated $\delta \approx 1/2$ behavior. Both layers have the same exponent when $A$ is small, as expected, while the precursor layer seems to grow with a larger exponent than the supernatant when $A$ increases.

The different $\delta$ exponents found in both geometries, rectangular and circular, may be attributed to the relationship between the length of the front and the reservoir size in both geometries. In the rectangular geometry both quantities are equal to the lateral size of the system $L_y$, defined as one of the parameters of the simulation. In contrast, in the circular geometry the length of the front (whose value is $L_f(t)=2 \pi \bar{h}(t)$) increases with time as the mean front position does, while the reservoir size remains constant. The time-decreasing ratio between the number of the cells in the reservoir and the number of cells in the front indicates that, in the circular geometry, it is more difficult for the driving to keep feeding the film with particles to maintain the expected growth.

\begin{figure}[t]
\centering
\includegraphics[width=0.45\textwidth]{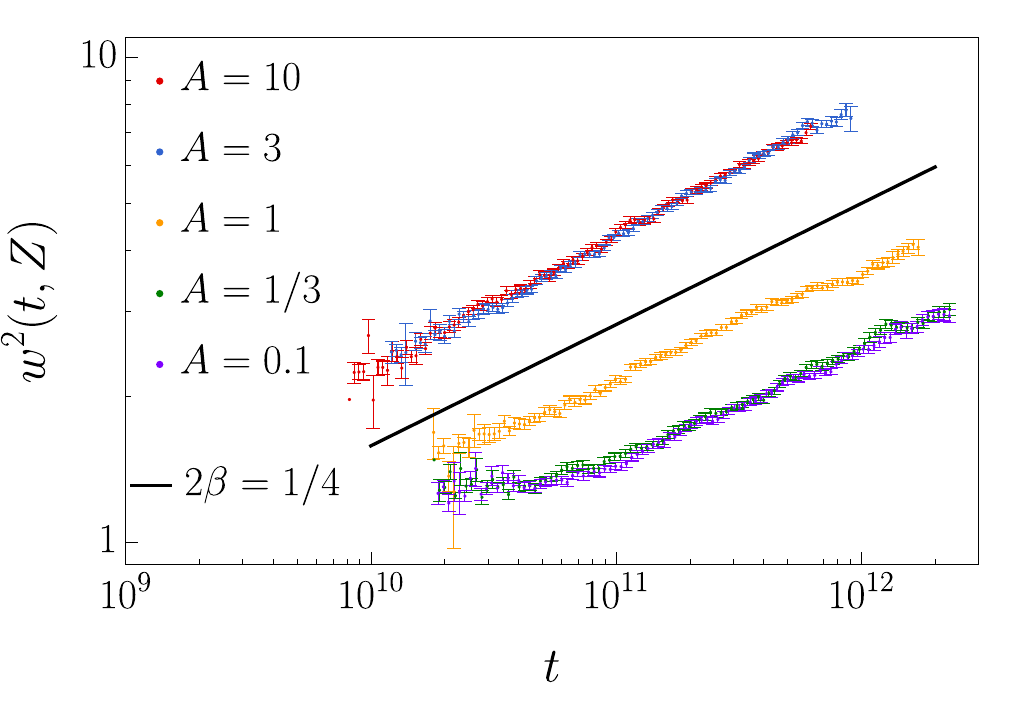}
\caption{(Color online) Squared roughness $w^2(t)$ as a function of time for $T=1/2$, $R_R=11$, $Z=1$, and several values of $A$, see legend. The solid black line corresponds to the reference scaling $w^2(t,Z) \sim t^{1/4}$.}
\label{fig:w2}
\end{figure}

With regard to the roughness, it scales as $w^2(t) \sim t^{2\beta}$, as expected. This behavior is shown in Fig.\ \ref{fig:w2} for a number of values of the Hamaker constant. No evidence of saturation to a steady-state value \cite{Barabasi1995,Krug1997} has been observed. Actually, as previously noted, steady-state saturation of the roughness was not expected in our system, given that the system size increases at a faster rate than the correlation length. Thus, the precursor film grows indefinitely and no finite equilibrium is reached \cite{Hocking1992}.

The value of $\beta$ depends upon the physical parameters $A$ and $T$, as demonstrated in Table \ref{tab:delta_precursor} (and in Table \ref{tab:delta_supernatant} in Appendix~\ref{extra}) for both layers. In Fig.\ \ref{fig:TablaExponentes} (bottom) we have plotted the dependence of the $\beta$ exponent as a function of $T$ for several values of $A$. At high temperatures (approximately $T > 1$), $\beta$ ranges between $1/4$ and $1/5$ and depends very little on the Hamaker constant $A$. At low temperatures $(T < 1)$, the growth exponent decreases until reaching $\beta \approx 1/10$. Once again, both layers exhibit the same exponent when $A$ is small, while the precursor layer exhibits a larger exponent when $A$ increases.

For conditions with very low temperatures and low Hamaker constants, the shape of the front is no longer circular. In these cases it is more appropriate to study its fluctuations locally. In particular, a new front width and a new $\beta_\Omega$ exponent can be computed from Eq.\ \eqref{eq:width_shape}. Table \ref{tab:beta_nueva_forma} shows the values of the $\beta_\Omega$ exponent computed following the aforementioned procedure. As expected, both $\beta$ and $\beta_\Omega$ exponents take on comparable values at high temperatures, as the local and global fluctuations are the same. Nevertheless, for very low temperatures ($T=1/3$) the exponent computed through this procedure is significantly greater.

Overall, the values of the exponents already indicate a non-trivial dependence with temperature and a much weaker dependence with the Hamaker constant. Thus, as observed in band geometry \cite{Marcos2022}, two main scaling regimes seem to exist, at low and at high temperatures, with $T$-dependent exponents for intermediate values of $T$. As will be discussed below, further exponent estimates confirm this picture.


\begin{figure}[t]
\centering
\includegraphics[width=0.45\textwidth]{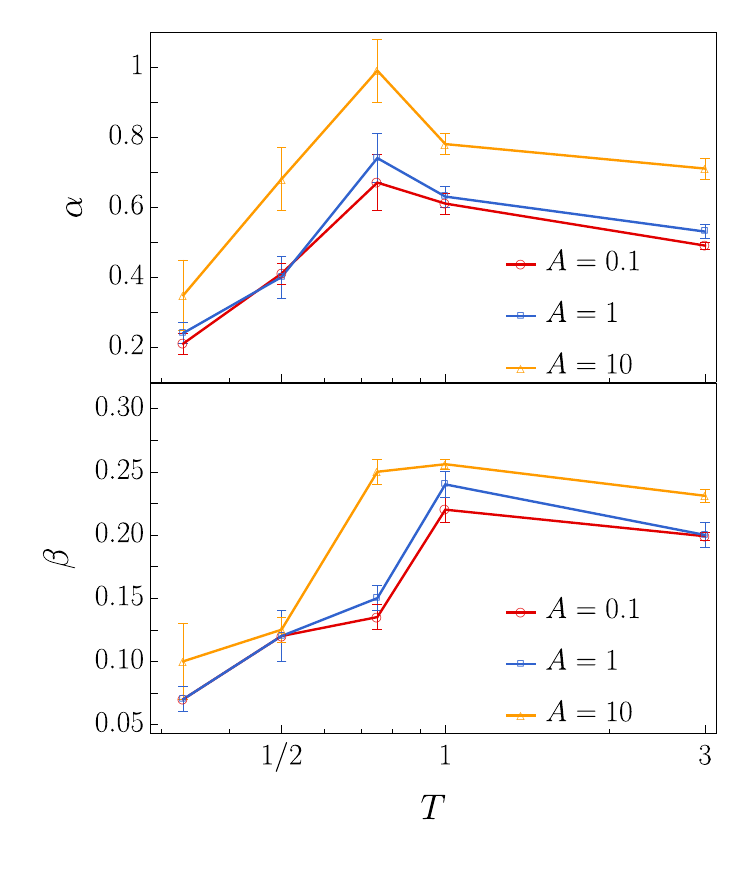}
\caption{(Color online) Values of $\alpha$ (top) and $\beta$ (bottom) for the precursor film (taken from Tables \ref{tab:delta_precursor} and \ref{tab:z_precursor}) vs $T$ for $A=0.1$ (red circles), $A=1$ (blue squares), and $A=10$ (orange triangles). Lines are guides to the eye.}
\label{fig:TablaExponentes}
\end{figure}

\begin{table*}[t]
\begin{ruledtabular}
\begin{tabular}{|c|c|c|c|c|c|}
\diagbox[width=2.5em]{$A$}{$T$} & 3 & 1 & 3/4 & 1/2 & 1/3\\ \hline
 \hline
        10 & 0.48(1) & 0.512(8) & 0.50(1) & 0.25(2) & 0.20(3) \\ \hline
        3 & 0.436(6) & 0.50(1) & 0.50(1) & 0.22(3) & 0.21(3) \\ \hline
        1 & 0.42(2) & 0.49(1) & 0.43(2) & 0.22(3) & 0.17(2) \\ \hline
        1/3 & 0.415(8) & 0.44(1) & 0.25(2) & 0.21(3) & 0.22(3)\\ \hline
        0.1 & 0.416(7) & 0.44(1) & 0.27(2) & 0.23(3) & 0.24(3)\\
\end{tabular}
\end{ruledtabular}
\caption{Values of the exponent $2\beta_{\Omega}$ for the precursor layer, computed using Eq.\ \eqref{eq:width_shape} for all the conditions studied.}
\label{tab:beta_nueva_forma}
\end{table*}

\subsection{Height-difference correlation function: computation of $\alpha$ and $z$ exponents}\label{sec:hdcf}

\begin{figure}[t]
\centering
\includegraphics[width=0.45\textwidth]{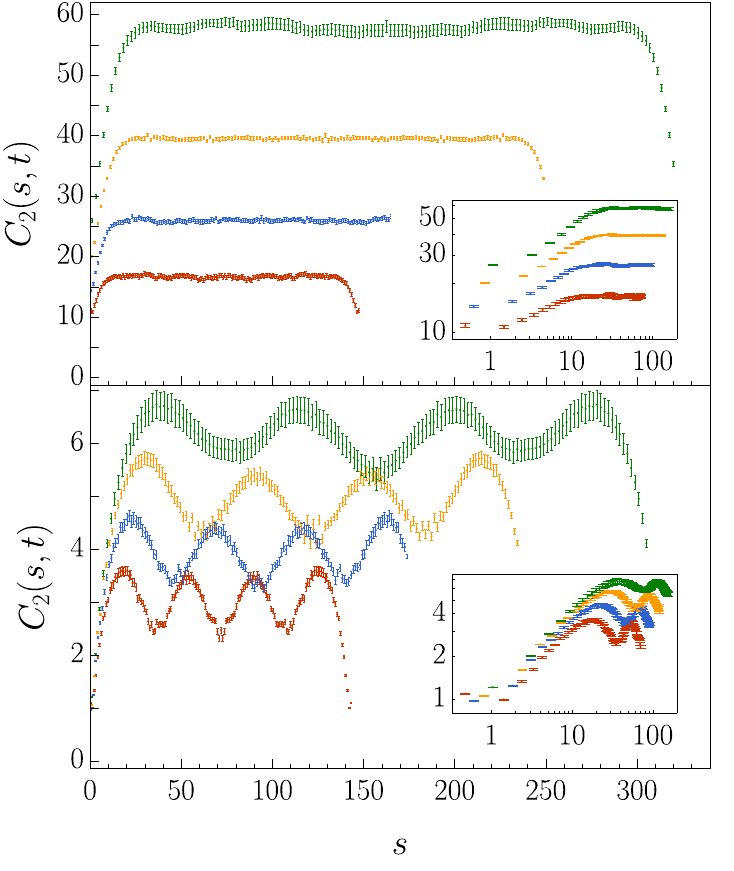}
\caption{(Color online) Height-difference correlation function $C_2(s,t)$ as a function of $s$ for the following time-boxes: $\{40,60,80,100\}$. Conditions are: $T=3$, $A=1$ (top), and $T=1/2$, $A=0.1$ (bottom). Inset: Log-log plot of the same function for $s$ between $0$ and $L_f(t)/2$.}
\label{fig:CorrelacionTotal}
\end{figure}

\begin{figure}[t]
\centering
\includegraphics[width=0.45\textwidth]{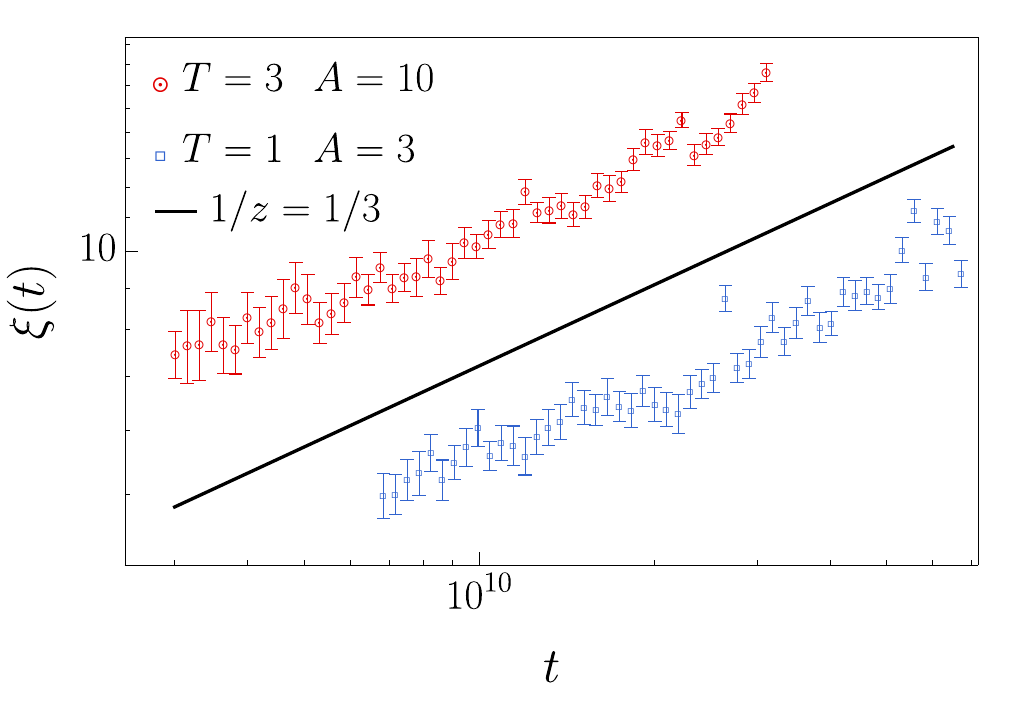}
\caption{(Color online)  Correlation length $\xi(t)$ for $T=3$, $A=10$ (red circles) and $T=1$, $A=3$ (blue squares) as functions of time. As a visual reference, the solid black line corresponds to $\xi(t)\sim t^{1/z}$, with $1/z=1/3$.}
\label{fig:xi}
\end{figure}

\begin{figure}[t]
\centering
\includegraphics[width=0.45\textwidth]{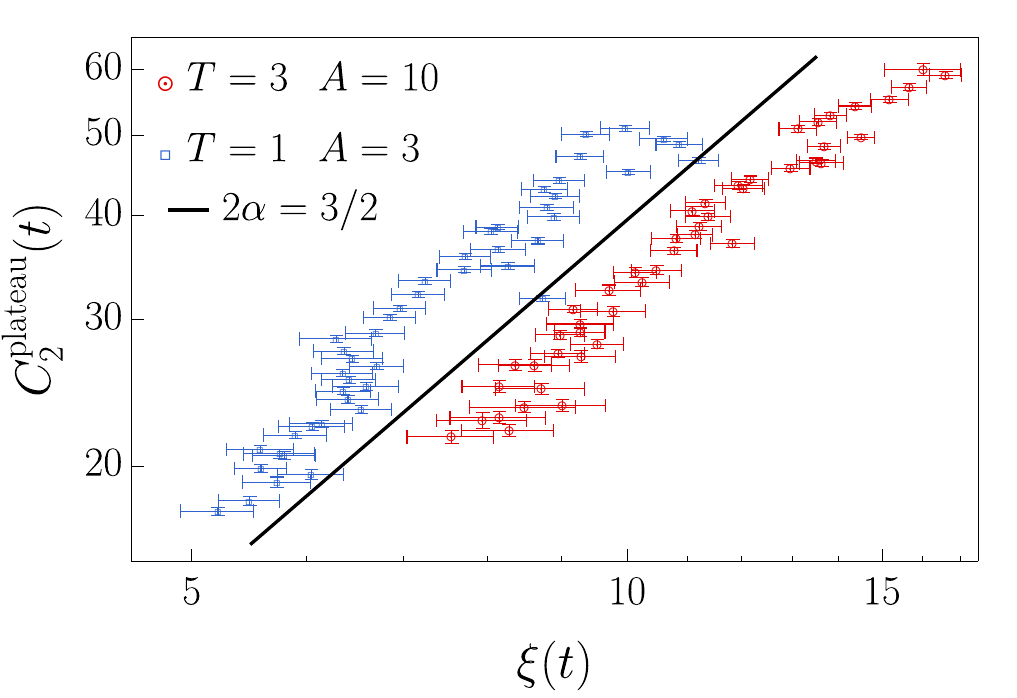}
\caption{(Color online) Height-difference correlation function $C_2^\mathrm{plateau}(t)$ versus $\xi(t)$  for $T=3$, $A=10$ (red circles) and $T=1$, $A=3$ (blue squares) at different times. As a visual reference, the solid black line corresponds to $C_2^\mathrm{plateau}(t)\sim \xi(t)^{2\alpha}$, with $2\alpha=3/2$.}
\label{fig:plateau}
\end{figure}

Figure \ref{fig:CorrelacionTotal} shows the height-difference correlation function at several times for two representative parameter conditions at high and low temperatures. As outlined in Sec.\ \ref{observables}, the correlation length $\xi(t)$ can be estimated from the plateau of the $C_2(s,t)$ curves at sufficiently large values of $s$, for a given value of $a$, when the height-difference correlation function reaches a plateau. As the values of the exponents are independent of the precise value of the parameter $a$~\cite{Barreales2020,Marcos2022}, in this work we used $a=0.9$. According to Eq.\ \eqref{eq:correlation_length}, the log-log plots of the correlation length as a function of time should fit a straight line whose slope is $1/z$. Figure \ref{fig:xi} shows log-log plots of $\xi(t)$ vs. $t$ for the precursor layer, calculated for two conditions with $1/z\sim 1/3$. On the other hand, Eq.\ \eqref{eq:corr_length} yields $C_2(s,t)=\xi^{2\alpha}(t)$ for $s \gg \xi(t)$. Thus, the $\alpha$ exponent may be calculated from the slope of the best-fit lines in a log-log plot of $C_2(s,t)$ versus $\xi(t)$. For the sake of simplicity, the correlation function has been evaluated at the center of the plateau of the height-difference correlation function. Figure \ref{fig:plateau} shows $C_2^\mathrm{plateau}(t)$ against $\xi(t)$ for the precursor layer and two conditions; the straight line with $2 \alpha = 3/2$ is plotted as reference.

When the height-difference correlation function does not reach a plateau, the same computations can be carried out to compute the exponents $1/z$ and $2\alpha$, but redefining the way the correlation length and plateau are calculated, as detailed in Sec.\ \ref{observables} (see also Appendix \ref{parabola} for details).

The complete set of $1/z$ and $2\alpha$ exponents for the precursor and supernatant layers are provided in Table \ref{tab:z_precursor} (and Table \ref{tab:z_supernatant} in Appendix~\ref{details}). From these data, one may easily verify that the expected scaling relation $\alpha=\beta z$ holds well for the majority of the conditions under study. A representative selection of these results for the $\alpha$ and $\beta$ exponents is presented graphically in Fig.\ \ref{fig:TablaExponentes}. As anticipated above with regard to the dependence of $\beta$ on temperature, the dependence of $\alpha$ and $z$ with temperature is of importance as well, and similarly suggests a transition from a low-temperature to a high-temperature regime, with $T$-dependent exponents for $T<1$. Furthermore, the dependence of $\alpha$ and $z$ with the Hamaker constant seems more significant than in band geometry, especially at low temperature. 
Moreover, in the top panel of Fig.\ \ref{fig:TablaExponentes} $\alpha$ shows a peak for intermediate temperatures. This behavior could be due to uncertainties being greater in the transition zone between the high and low temperature regimes and is also observed in the system studied in the band geometry \cite{Marcos2022}.

The exponents measured in our simulations —-$\alpha$, $\beta$, and $z$—-, which characterize the space-time fluctuations of a growing surface or front, in principle can be measured experimentally \cite{Barabasi1995,Takeuchi2018}. For example, 
in the case of reactive spreading of fluid films [Hg droplets on metal-on-glass (with Au or Ag) substrates], such type of measurements are reported (for the band-geometry case) e.g.\ in Refs.\ \cite{Beer2008,Yin2009,Harel2017} and other therein. However, we are not aware of similar measurements for our present case of non-reactive fluid spreading. The results from our model which seem most relevant for comparison with experiments correspond to the parameter regime where the precursor film is present, specifically at low \( T \) and high \( A \). Notably, under this parameter choice, our model system exhibits growth behavior closely matching the experimental findings, at least in terms of the \( R(t) \sim t^{1/2} \) relation.

\begin{table*}[t]
\begin{ruledtabular}
\begin{tabular}{|c|c|c|c|c|c|c|c|c|c|c|}
\multirow{2}{*}{\diagbox[width=2.5em]{$A$}{$T$}} & \multicolumn{2}{c|}{3}  & \multicolumn{2}{c|}{1}  & \multicolumn{2}{c|}{3/4}  & \multicolumn{2}{c|}{1/2} & \multicolumn{2}{c|}{1/3} \\  \cline{2-11}

 & $1/z$ & $2\alpha$&$1/z$ & $2\alpha$& $1/z$ & $2\alpha$& $1/z$ & $2\alpha$& $1/z$ & $2\alpha$ \\ \hline
 \hline
         10 & 0.33(4)& 1.42(6)& 0.31(1)& 1.56(6) & 0.21(2)& 2.0(2)& 0.18(2)& 1.4(2)& 0.26(5)& 0.7(2)\\ \hline
        3 & 0.38(1)& 1.13(3)& 0.32(1)& 1.48(7) & 0.21(2)& 2.0(2)& 0.24(2)& 1.0(1)& 0.29(6)& 0.7(2)\\ \hline
        1 & 0.37(1)& 1.07(3)& 0.35(2)& 1.27(7) & 0.25(2)& 1.5(1)& 0.20(3)& 0.8(1)& \bf{ 0.28(1)}& \bf{ 0.48(6)}\\ \hline
        1/3 & 0.38(1)& 1.02(3)& 0.28(1)& 1.43(7) & 0.18(2)& 1.3(2)& \bf{ 0.26(2)}& \bf{ 0.80(6)} &\bf{ 0.28(1)}& \bf{ 0.45(6)}\\ \hline
        0.1 & 0.40(1)& 0.98(3)& 0.31(2)& 1.22(6) & 0.19(2)& 1.3(2)& \bf{ 0.25(2)}& \bf{ 0.83(5)}& \bf{ 0.28(1)}& \bf{ 0.42(6)}\\
\end{tabular}
\end{ruledtabular}
\caption{Values of the exponents $1/z$ and $2\alpha$, for the precursor layer, for all the conditions under study. The values calculated approximating the peak as a parabola appear in bold.}
\label{tab:z_precursor}
\end{table*}

\subsection{Shape of the height-difference correlation function at low temperature}\label{shape}

Figure \ref{fig:CorrelacionTotal} (bottom) shows a low-temperature height-difference correlation function at several times; the presence of several maxima and minima is evident. Besides, as illustrated in Fig.\ \ref{fig:morfologia}, the films, particularly the precursor film, exhibit a shape that resembles a square at low temperature and low Hamaker constant conditions. These two facts are related to each other. In contrast with the expected plateau, the height-difference correlation function at low temperature exhibits four peaks (and three local minima), indicating that points separated by arc-length distances of $\pi \bar{h}/2$, $\pi \bar{h}$, and $3\pi \bar{h}/2$ are less correlated than those separated by larger or smaller arc-lengths. It should be noted that the height-difference correlation function of a perfect square (not shown here) exhibits four perfectly symmetric peaks and three local minima that go to zero at the same locations as those of the local correlation minima observed in our case.

Thus, the data indicate that the low-temperature films take on a shape that is intermediate between a square and a circle. To test this hypothesis, we have computed the average shape of the film for a single low-temperature condition across multiple runs. The result is shown in Fig.\ \ref{fig:cuadrado} of Appendix \ref{extra}, where a square-like rounded shape can be clearly seen. It is noticeable that the transition between always-occupied and always-empty cells is abrupt, resulting into a film that resembles a square with rounded corners.

The reason why our system adopts this form can be explained considering how it evolves in time. Indeed, our system implements the same (Kawasaki) dynamics as the Conserved Order Parameter (COP) Ising model, thereby displaying a similar behavior \cite{Newman1999}. In the COP Ising model, the shape of the domain at very low temperature (approximately $T/T_c=0.25$, with $T_c\approx 2.27$) resembles a square while, at higher temperatures, it becomes rounded (see e.g.\ Fig.\ 5.4  of Ref.\ \cite{Newman1999}). This phenomenon occurs because the system minimizes its energy by minimizing the perimeter of the domain. The dynamics of our system resembles that of the COP Ising model, since they both are defined in a regular lattice and they share the Kawasaki exchange rules. However, there are two notable distinctions. The first one is the second term in our Hamiltonian, Eq.\ \eqref{eq:energy}, i.e. the interaction with the substrate. The second one is the existence of a reservoir that continually adds particles to the system. Nevertheless, these differences are not relevant in the context of the film morphology analysis. Indeed, the main effect of the interaction with the substrate is to favor the growth of the precursor film, but this interaction does not change the energy of the particles in the same layer and therefore it is not relevant when studying the shape that the films adopt. On the other hand, the fact that the reservoir adds particles to the films is also irrelevant for long times, since the rate at which the particles reach the front slows down with time. For these long times the algorithm performs numerous steps without change in the number of particles in the system. Therefore, the energy of the system is minimized by reducing its perimeter, as observed in the COP Ising model, resulting in the adoption of a square shape with rounded corners. This characteristic shape arises from the simplicity of the model and the choice of the lattice. Other authors who have similarly studied fluid droplets through kMC simulations of discrete models have found that this rectangular shape disappears, provided interactions beyond the first neighbors are allowed for \cite{Areshi2019,Chalmers2017}. However, employing those in our case would severely hamper our computational capacity to assess scaling behavior, that requires sufficiently long times and large system sizes. In addition, other authors who have studied the Ising model with Kawasaki dynamics in a hexagonal lattice in the limit of vanishing temperature have found that the equilibrium state of the system is a hexagon \cite{Baldassarri2023}. This characteristic shape should not be considered as a physically meaningful result, but as a feature of the model itself.

\subsection{Anomalous scaling of the height-difference correlation function}

\begin{table*}[t]
\begin{ruledtabular}
\begin{tabular}{|c|c|c|c|c|c|}
\diagbox[width=2.5em]{$A$}{$T$} & 3 & 1 & 3/4 & 1/2 & 1/3\\ \hline
 \hline
        10 & 1.18(6)& 1.44(6)& 1.9(2)& 0.9(2)& 0.0(2)\\ \hline
        3 & 0.90(3)& 1.35(7)& 1.9(2)& 0.7(1)& 0.0(2)\\ \hline
        1 & 0.84(3)& 1.15(6)& 1.4(1)& 0.4(1)& $-0.66(6)$ \\ \hline
        1/3 & 0.81(3)& 1.33(7)& 1.1(2)& 0.30(6)& $-0.62(6)$ \\  \hline
        0.1 & 0.78(3)& 1.11(6)& 1.1(2)& 0.31(5)& $-0.73(6)$ \\ 
\end{tabular}
\end{ruledtabular}
\caption{Value of the exponent $2\alpha'$, for the precursor layer, for all the conditions studied.}
\label{tab:dap_precursor}
\end{table*}

\begin{figure}[t]
\centering
\includegraphics[width=0.45\textwidth]{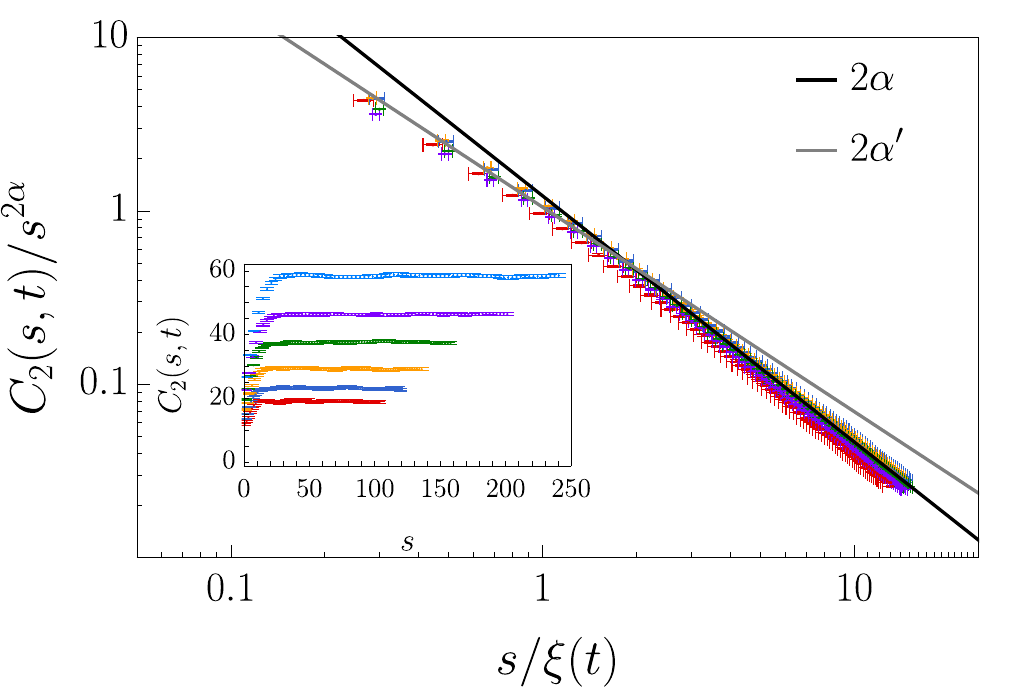}
\caption{(Color online) Data collapse of the height-difference correlation function obtained for different values of time, for $J = 1$, $Z=1$, $T = 3$, and $A = 10$. The curve onto which collapse occurs is the function $g(s/\xi(t))$ of Eq.\ \eqref{eq:corr_length}, with the solid black line representing the theoretical behavior for large argument, $g(u)\sim u^{-2\alpha}$ with $2\alpha=1.42$, and the solid gray line representing the behavior for small argument, $g(u)\sim u^{-2\alpha'}$  with $2\alpha'=1.18$. Inset: height-difference correlation function as a function of $s$ for times increasing from 50 to 100 bottom to top at regular intervals.}
\label{fig:reescalada}
\end{figure}

The fact that the $C_2(s,t)$ curves obtained for different times shift systematically with time and do not overlap for $s<\xi(t)$, as shown in Fig.\ \ref{fig:CorrelacionTotal}, is a clear indication of anomalous scaling behavior \cite{Lopez1997}. The existence of intrinsic anomalous scaling stems from the fact that $\alpha_{\rm loc} < \alpha$, there being two independent roughness exponents. This time shift is also unambiguously shown in the main panel of Fig.\ \ref{fig:reescalada}, which adds a consistent data collapse of the height-difference correlation function according to Eq.\ \eqref{eq:corr_length} for a representative parameter choice. 



In the presence of anomalous scaling, the height-difference correlation function does not follow the standard FV scaling, according to which $g(u)$ should be $u$-independent at small arguments $u \ll 1$, see Eq.~\eqref{eq:corr_length}. Rather, $g(u) \sim u^{-2\alpha'}$ for $u \ll 1$, with $\alpha' = \alpha - \alpha_\text{loc}$. We have computed the value of $2\alpha'$ by fitting the re-scaled height-difference correlation function $C_2(s,t)/s^{2\alpha}$ vs $s/\xi(t)$ for $s/\xi(t)<1$ and several times. Table \ref{tab:dap_precursor} (and Table \ref{tab:dap_supernatant} in Appendix~\ref{details}) lists the $2\alpha'$ values thus obtained for the precursor and supernatant layers, respectively, for several choices of $A$ and $T$. According to these tables, the exponents depend heavily on the parameter conditions. In addition, although the anomalous shift of the height-difference correlation function curves with increasing time shown in the inset of Fig.\ \ref{fig:reescalada} could be induced by a mere large roughness exponent, the collapse of the same data in Fig.\ \ref{fig:reescalada} with $\alpha'\neq 0$ unambiguously identifies the origin of this behavior, rather, as intrinsic anomalous scaling. Note that there is a couple of cases ($T=1/3$ and $A=10$ and 3) for which the relation $\alpha\ne\alpha'$ does not hold.

\begin{figure}[t]
\centering
\includegraphics[width=0.45\textwidth]{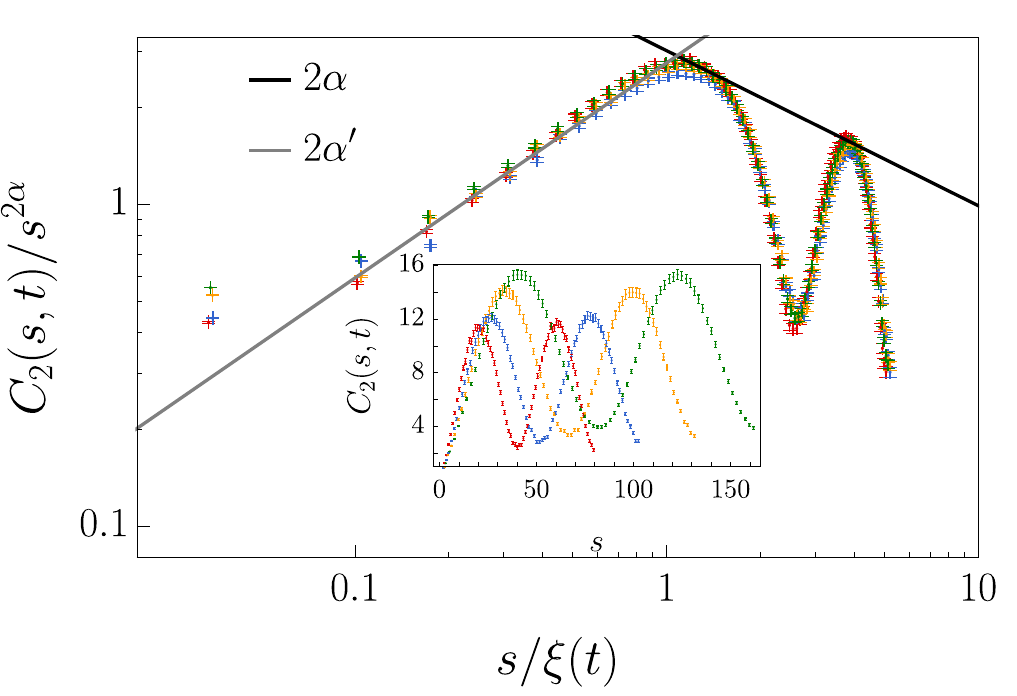}
\caption{(Color online) Data collapse according to Eq.\ \eqref{eq:h}, for the height-difference correlation function obtained for different values of time, for $Z=1$, $T = 1/3$, and $A = 1$. The solid black line corresponds to $g(u)\sim u^{-2\alpha}$ with $2\alpha=0.48$, and the solid gray line corresponds to $g(u)\sim u^{-2\alpha'}$  with $2\alpha'=-0.66$. Inset: height-difference correlation function as a function of $s$ for times increasing from 55 to 100 bottom to top at regular intervals.}
\label{fig:reescaladaTbaja}
\end{figure}

For those conditions where the height-difference correlation function oscillates, notably it is still possible to perform a data collapse analogous to Eq.\ \eqref{eq:corr_length}. An illustrative example of such a collapse is shown in the main panel of Fig.\ \ref{fig:reescaladaTbaja}. In these low-temperature cases, however, the specific scaling function $h(u)$ to which the height-difference correlation data collapse differs from that of Eq.\ \eqref{eq:corr_length} and Fig.\ \ref{fig:reescalada}. Specifically, Eq.\ \eqref{eq:corr_length} is modified into
\begin{equation}
    C_2(s,t)=s^{2\alpha} h(s/\xi(t)) \,,
    \label{eq:h}
\end{equation}
where $h(u) \sim u^{-2\alpha'}$ for $u\ll 1$; for $u\gg 1$, the function $h(u)$ oscillates with an amplitude that decays as $1/u^{2\alpha}$, see Fig.\ \ref{fig:reescaladaTbaja}; note also
that in the figure $2\alpha'$ takes a negative value.

\subsection{Additional universal properties of the fronts: probability distribution function}

Recent developments on surface kinetic roughening, particularly in the context of KPZ scaling, have demonstrated that the universal behavior extends beyond the values of the critical exponents for many important universality classes. Specifically, by normalizing the fluctuations of the front around its mean by their time-dependent amplitude as
\begin{equation}
	\label{eq:fluctuations}
	\chi_i(t) = \frac {h_i(t) - \bar{h}(t)}{t^{\beta}} ,
\end{equation}
the probability density function (PDF) of these $\chi$ random variables becomes time-independent and is shared by all members of the universality class \cite{Kriecherbauer2010,HalpinHealy2015,Carrasco2016,Takeuchi2018,Carrasco2019}.

\begin{figure}[t]
\centering
\includegraphics[width=0.45\textwidth]{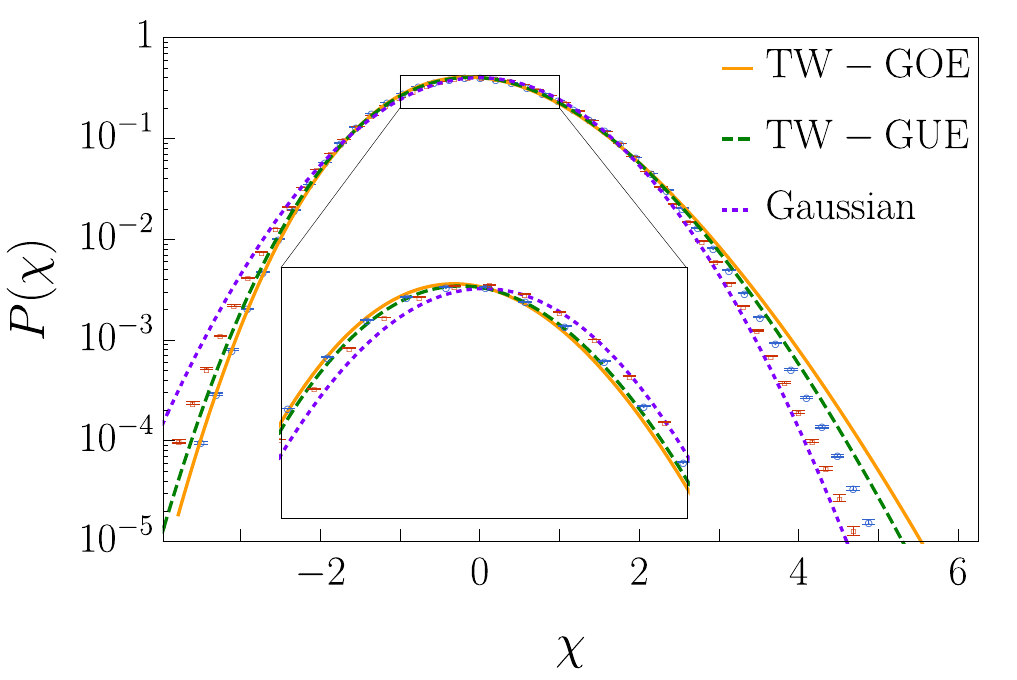}
\caption{(Color online) Fluctuation histograms of the variable $\chi$ calculated according to Eq.\ \eqref{eq:fluctuations} for $T = 1$, $A=1$ (blue circles) and $T = 3$, $A=10$ (red squares). The solid orange line and the green dashed line correspond to the GOE and GUE Tracy-Widom distributions, respectively. The dotted purple line correspond to a Gaussian distribution. Inset: zoom for small $\chi$.}
\label{fig:histograma}
\end{figure}

In a previous work \cite{Marcos2022} we found that in this same system, but in a band geometry with periodic boundary conditions, the front fluctuations follow the same distribution as in the KPZ universality class, i.e.\ the TW-GOE PDF. 
Figure \ref{fig:histograma} shows both, the TW-GOE PDF together with the TW-GUE PDF that holds for the KPZ universality class on a circular geometry, along with the Gaussian distribution, and data for the two relevant conditions from our numerical simulations. The agreement with the TW-GUE PDF is notable, specially considering that the exponents of the system are not those of the KPZ universality class. In our previous work \cite{Marcos2022}, we demonstrated that the agreement improved with increasing the system size, suggesting that the inaccuracies found were due to finite-size effects. In our present case, the length of the film $L_f(t)=2\pi\bar{h}(t)$ is not a parameter that can be controlled, but rather a function that grows with time.
We have found the best agreement for high temperatures ($T \ge 3/4$), while some discrepancies were seen for smaller temperatures. We also report a worse agreement for smaller Hamaker constants, although this parameter is somewhat less relevant than temperature. In turn, for low temperatures the tail of the distribution seems to get closer to the Gaussian than to the TW-GOE and TW-GUE PDFs.

\begin{figure}[t]
\centering
\includegraphics[width=0.45\textwidth]{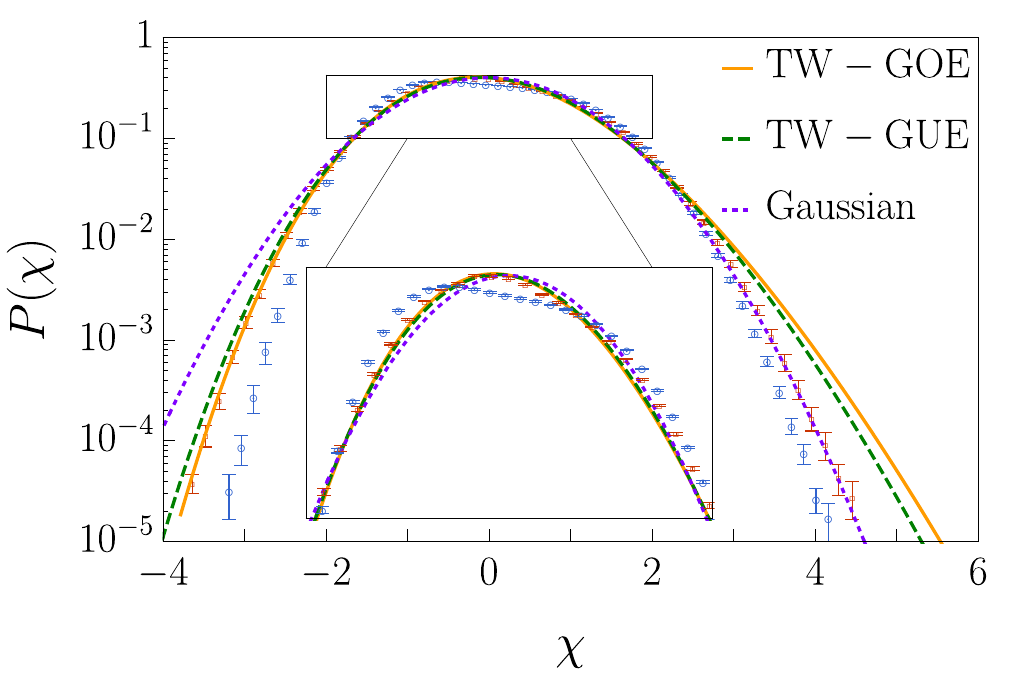}
\caption{(Color online) Fluctuation histograms of the variable $\chi$ calculated for $T = 1/3$, $A=1$ according to Eq.\ \eqref{eq:fluctuations}, i.e. $\chi$, (shown in blue circles) and according to Eq.\ \eqref{eq:Chi_local}, i.e.  $\chi_\Omega$, (shown in red squares). The solid orange and the green dashed lines correspond to the GOE and GUE Tracy-Widom distributions, respectively. The dotted purple line correspond to a Gaussian distribution. In each case, the growth exponent used was the one calculated with each method, i.e. the one appearing in Table \ref{tab:delta_precursor}, $\beta$, for the first case and the one appearing in Table \ref{tab:beta_nueva_forma}, $\beta_\Omega$, for the second case. Inset: zoom for small $\chi$ and $\chi_\Omega$.}
\label{fig:histograma_Tbaja}
\end{figure}

In order to clarify the fluctuation statistics at low $T$, we have also assessed the statistics of front fluctuations by measuring the deviations from the local average front, as defined in Eq.\ \ref{eq:width_shape}. This approach is expected to improve the results in systems which develop characteristic shapes \cite{Domenech2024}, as in our case for low temperatures. Thus, we define the random variables
\begin{equation}
	\label{eq:Chi_local}
	{\chi_{\Omega}}_i(t)=\frac{h_i(t)-\bar{h}_\Omega(t,Z)}{t^{\beta_{\Omega}}},
\end{equation}
where $\beta_\Omega$ quantifies the time increase of the local roughness $w_{\Omega}(t)$ defined in Eq.\ \eqref{eq:width_shape}. Figure \ref{fig:histograma_Tbaja} shows the fluctuation PDFs  obtained for a particular condition in which the shape is relevant (namely $T=1/3$ and $A=1$), by measuring fluctuations globally [i.e., following Eq.\ \eqref{eq:fluctuations}] and locally [i.e., following Eq.\ \eqref{eq:Chi_local}]. As illustrated in this figure, this methodology for quantifying local fluctuations is indeed particularly well-suited to scenarios where the film shape deviates from a circular configuration \cite{Domenech2024}.

In addition to the fluctuation PDF and as a confirmation of our results, we have also directly computed its third and fourth order cumulants, namely the skewness and excess kurtosis, whose values are analytically known for the Gaussian, TW-GOE, and TW-GUE distributions appearing in Figs.\ \ref{fig:histograma} and \ref{fig:histograma_Tbaja} \cite{Kriecherbauer2010,Takeuchi2011}.\footnote{Specifically, for the cases shown in Fig.\ \ref{fig:histograma}, we found that $S=0.207(2)$, $K=0.063(4)$ for $T=1$ and $A=1$, while $S=0.086(2)$, $K=0.061(3)$ for $T=3$ and $A=10$. For reference the precise skewness and excess kurtosis values are $S=0.29346452408$ and $K=0.1652429384$ for the TW-GOE and $S=0.224084203610$ and $K=0.0934480876$ for the TW-GUE \cite{Bornemann2010}.} Writing the local height fluctuations as $\delta h_i(t)=h_i(t)-\bar{h}(t)$, we define the skewness and the excess kurtosis as $S~=~\langle \delta h_i(t)^3 \rangle / \langle \delta h_i(t)^2 \rangle^{3/2}$ and $K=\langle \delta h_i(t)^4 \rangle / \langle \delta h_i(t)^2 \rangle^{2}-3$, respectively. 
In general we have found that $0<S<1/3$ and $0<K<0.25$, except in cases where the cumulants have been computed by subtracting the shape ($T=1/3$ and $A\le 1$), where $K\approx -0.2$. This can be perceived from the tails of the distribution in Fig.\ \ref{fig:histograma_Tbaja}.

\section{Summary and Conclusions}\label{sec:concl}

In summary, we have investigated through comprehensive kMC simulations the spatiotemporal dynamics of the fronts of circular liquid droplets spreading on flat substrates. To this end, we have used a discrete model of the system based on the Ising lattice gas, examining its behavior in relation to parameters such as the Hamaker constant (wettability) and temperature. This model had been previously studied in a band geometry in Ref.\ \cite{Marcos2022}. We have examined standard morphological observables, including the mean front position and its roughness, across a wide range of model parameters. Additionally, we conducted a systematic analysis of the two-point correlation functions and evaluated the probability distribution function of front fluctuations. The main findings for the discrete lattice gas model can be summarized as follows:

\begin{itemize}
\item The $\delta\approx 1/2$ value of the exponent characterizing the mean position of the front of the precursor film seems to be obtained only for the most realistic conditions for the film (i.e., low temperature and high Hamaker constant). In general, while the values we obtain for $\delta$ on a circular geometry are somewhat smaller than their counterparts on band geometry, they are still much larger than the classical Tanner values characteristic of the spreading of macroscopic droplets.
\item The critical exponents $\alpha$, $\beta$, and $z$ depend more strongly on temperature than on the Hamaker constant.
\item The values of the critical exponents show a transition from a low-temperature to a high-temperature regime, where they become largely $T$-independent. Although slight quantitative differences exist, these behaviors are very similar to those found for a band geometry.
\item The front exhibits intrinsic anomalous scaling, regardless of parameter values, such that the roughness exponents describing front fluctuations at large ($\alpha$) and small ($\alpha_{\rm loc}$) length scales differ. While this result is again in line with the behavior found on a band geometry, we are not aware of previous assessments of intrinsic anomalous scaling for interfaces with an overall circular symmetry.
\item When the temperature and Hamaker constant are both relatively low, the average film shape deviates significantly from a circle to a square. This is clearly evident in both the shape of the films and the form of the two-point correlation functions for those conditions.
\item We have developed a consistent method to compute the correlation length $\xi(t)$ for those cases in which the height-difference correlation function does not reach a plateau.
\item In spite of the facts that the exponent values differ from those of the 1D KPZ universality class and that the dynamics scaling ansatz also differs, being here intrinsically anomalous rather than FV, the PDFs of the front fluctuations show a reasonable degree of agreement with those of the 1D KPZ universality class in a circular geometry, namely, the TW-GUE PDF.
\end{itemize}

Admittedly, there remain some quantitative, rather than qualitative, differences of the present results with respect to those obtained for the same model in a rectangular geometry \cite{Marcos2022}. It should be noted, however, that the definition of the front is different in both geometries. Namely, in the rectangular geometry we used a single-valued approximation for the front, while in the present work we have presented a more complex and better suited definition for the front of the expanding circular droplets. Another difference between the two geometries is that, in rectangular geometry, the front length $L_f(t)$ is always equal to the reservoir size while, in a circular geometry, $L_f(t)$ increases with time, while the reservoir size remains constant. This results into a slower growth rate which challenges the study of this system in this geometry. At any rate, we believe that the combination of Ref.\ \cite{Marcos2022} and our present results advocates for a well-defined universality class for this type of film spreading processes, which combines intrinsic anomalous scaling with $T$-dependent exponents, and the dependence of interface geometry (through the subclass for the statistics of front fluctuations) that can be expected from 1D KPZ-related interfaces.

One of the most significant implications of anomalous scaling is that, unlike the standard Family-Vicsek case, where a single roughness exponent characterizes the fluctuations of the front morphology, this scenario involves two distinct roughness exponents governing local and global behaviors. Consequently, a thorough evaluation of kinetic roughening in experimental films necessitates measuring both exponents, for instance, using the $C_2(r,t)$ correlation function, as done in our analysis of simulation results. More broadly, the ability to clearly identify the kinetic roughening universality class in experiments can aid in validating theoretical models or descriptions of the underlying physical process.

\begin{acknowledgments}
This work was partially supported by Ministerio de Ciencia, Innovaci\'on y Universidades (Spain), Agencia Estatal de Investigaci\'on (AEI, Spain, 10.13039/501100011033), and European Regional Development Fund (ERDF, A way of making Europe) through Grants No.\ PID2020-112936GB-I00, No.\ PGC2018-094763-B-I00, and No.\ PID2021-123969NB-I00, and by the Junta de Extremadura (Spain) and Fondo Europeo de Desarrollo Regional (FEDER, EU) through Grants No.\ GR21014 and No.\ IB20079. J.\ M.\ Marcos is grateful to the Spanish Ministerio de Universidades for a predoctoral fellowship No. FPU2021-01334. We have run our simulations in the computing facilities of the Instituto de Computaci\'{o}n Cient\'{\i}fica Avanzada de Extremadura (ICCAEx).
\end{acknowledgments}

\appendix
\section{Details} \label{details}

In this appendix we present three subsections that provide additional information relevant to understanding the main text. 
In the first subsection we present additional comments on how the reservoir was chosen. In the second subsection we detail how we calculate the correlation length when the height-difference correlation function present peaks. Finally, in the last subsection we present some additional tables and figures, to support some issues described in the main part of the manuscript.

\subsection{Size and shape of the reservoir}\label{reservoir}
The size of the reservoir must be chosen with care. A small value can result into the same issues observed in single-cell reservoirs, namely that after a few steps of the kMC algorithm, a significant number of particles become disconnected from the reservoir. Consequently, the majority of transitions fail to contribute to the growth of the films, leading to a markedly slow algorithm. Conversely, if the reservoir is excessively large, the front will require a considerable time to grow away from it, as it must traverse a larger area to reach the same distance. The reservoir size adopted here falls between these two extremes. 

A further crucial issue is the shape of the reservoir. Besides the selected circular reservoir that was used for the simulations, a number of tests were performed with alternative geometrical shapes, including a square and an hexagon. The results were comparable in all cases. The circular reservoir was chosen due to its simplicity and the ease with which the pertinent physical quantities of interest can be computed.

\subsection{Analysis of oscillating height-difference correlation functions}\label{parabola}

\begin{figure}[t!]
\centering
\includegraphics[width=0.45\textwidth]{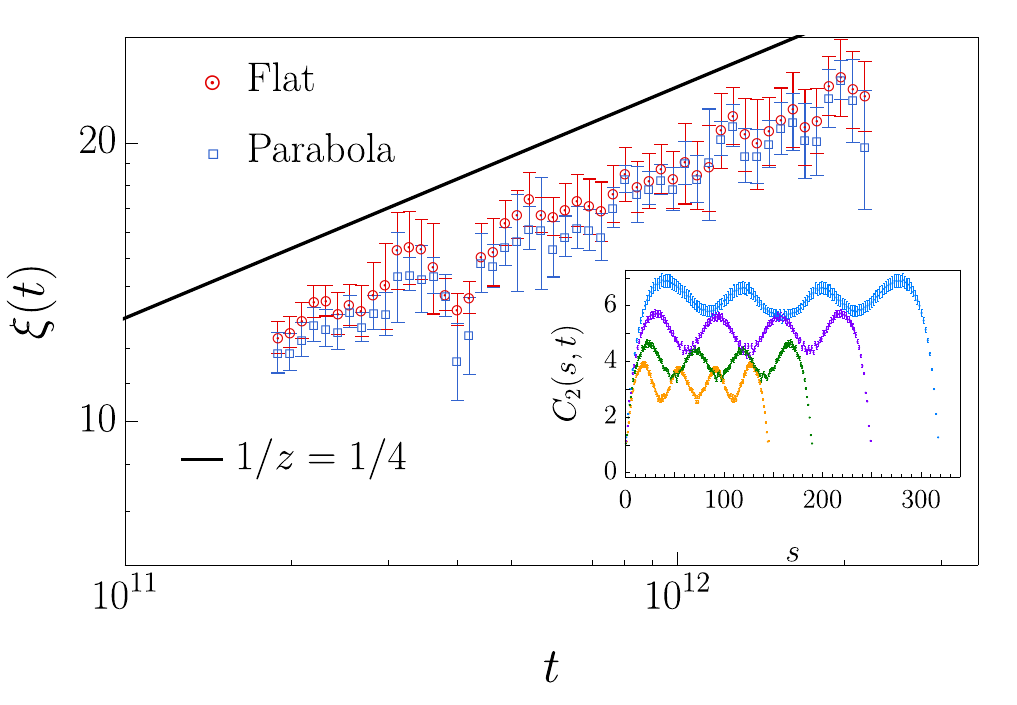}
\caption{(Color online) Correlation length $\xi(t)$ as a function of time, calculated using the flat method (red circles) and the parabola method (blue squares) for $T=1/2$ and $A=1/3$. As a visual reference, the solid black line corresponds to $\xi(t)\sim t^{1/z}$, with $1/z=1/4$. The quoted values of the exponents $1/z$ are $1/z=0.25(3)$ (flat) and $1/z=0.26(2)$ (parabola). Inset: height-difference correlation function as a function of $s$ for times increasing from 60 to 90, bottom to top, at regular intervals.}
\label{fig:equiv}
\end{figure}

Once the peak of the height-difference correlation function $C_2(s,t)$ has been fitted to a parabola $f(x)=a+bx+cx^2$, the plateau value is computed as the maximum value of the parabola, [i.e.\ $C(L/2,t)\equiv f_{\rm max}=a-b^2/(4c)$], and the correlation length as the point to the left of the parabola that reaches 0.9 of the plateau value, [i.e. $\xi(t)\equiv x_{0.9}=[-b+\sqrt{0.1(b^2-4ac)}]/(2c)$]. We have checked that, in the limits when both methods can be used ($T=1/2$ and low $A$, and $T=1/3$ and high $A$), the values of the correlation length and plateaus are the same. In Fig.~\ref{fig:equiv} we have plotted the correlation length $\xi(t)$ obtained by both methods. When these can be applied, we have selected those values with the smallest errors. In Tables \ref{tab:z_precursor} and \ref{tab:z_supernatant}, the conditions for which the parabolic approximation has been used appear in bold.

\subsection{Tables and additional Figures}\label{extra}

In this subsection we present some additional plots that are important to give some context to the main text. Additionally, Tables \ref{tab:delta_supernatant}, \ref{tab:z_supernatant}, and \ref{tab:dap_supernatant} report the values obtained for the exponents $\delta$, $\beta$, $\alpha$, $z$, and $\alpha'$ that correspond to the supernatant layer.

\begin{figure}[H]
\centering
\includegraphics[width=0.45\textwidth]{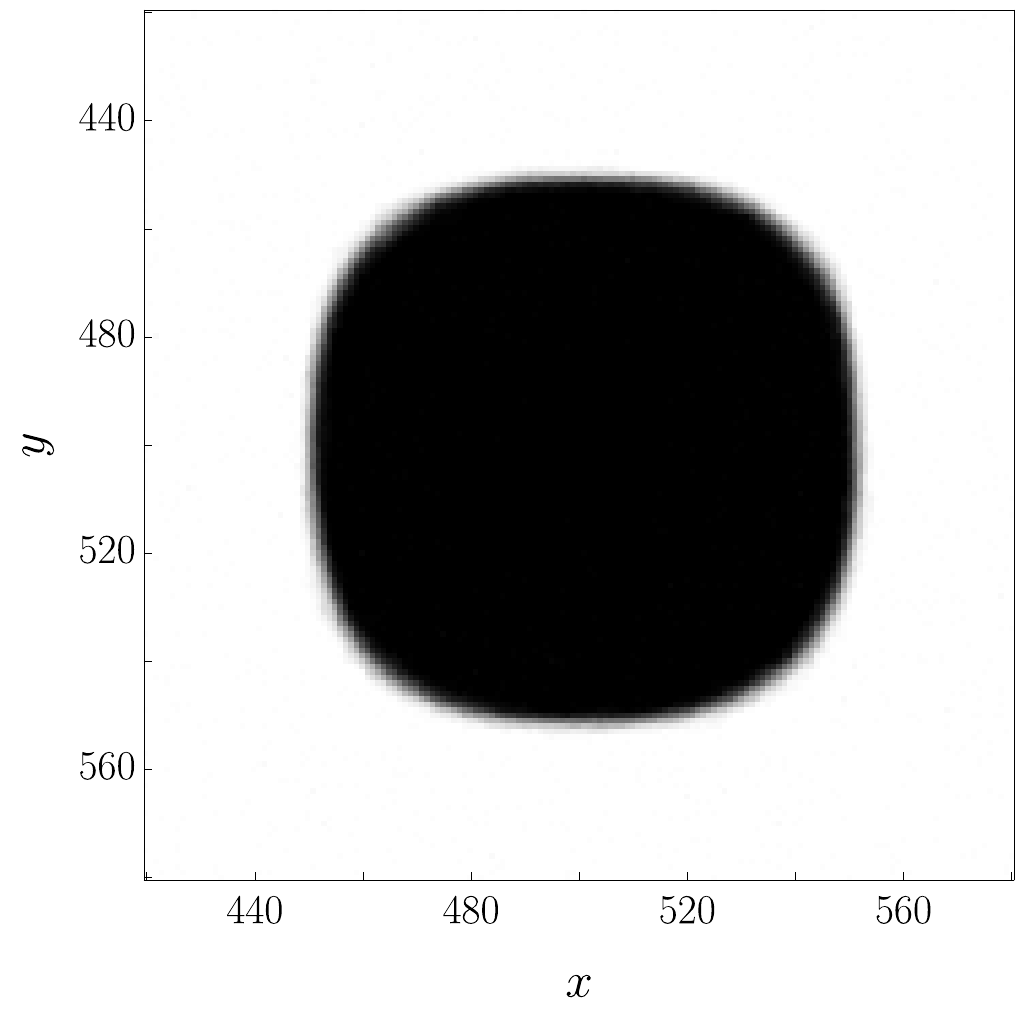}
\caption{Average of the last configurations measured for $T=1/3$, $A=1$, and $Z=1$ (i.e., the precursor film). The figure plots the gray level of the point density. In other words, a solid black cell (such as those belonging to the droplet reservoir at the center of the figure) indicates that the cell was occupied in all the runs. Conversely, a solid white cell is indicative that all the runs have this cell empty. Intermediate gray-level values represent varying degrees of density.}
\label{fig:cuadrado}
\end{figure}

Figure \ref{fig:cuadrado} illustrates an average of runs for a specific parameter condition ($T=1/3$ and $A=1$) which yields oscillations in the height-difference correlation function. The figure shows that the film shape under these conditions is that of a square with rounded corners.

\begin{figure}[H]
\centering
\includegraphics[width=0.45\textwidth]{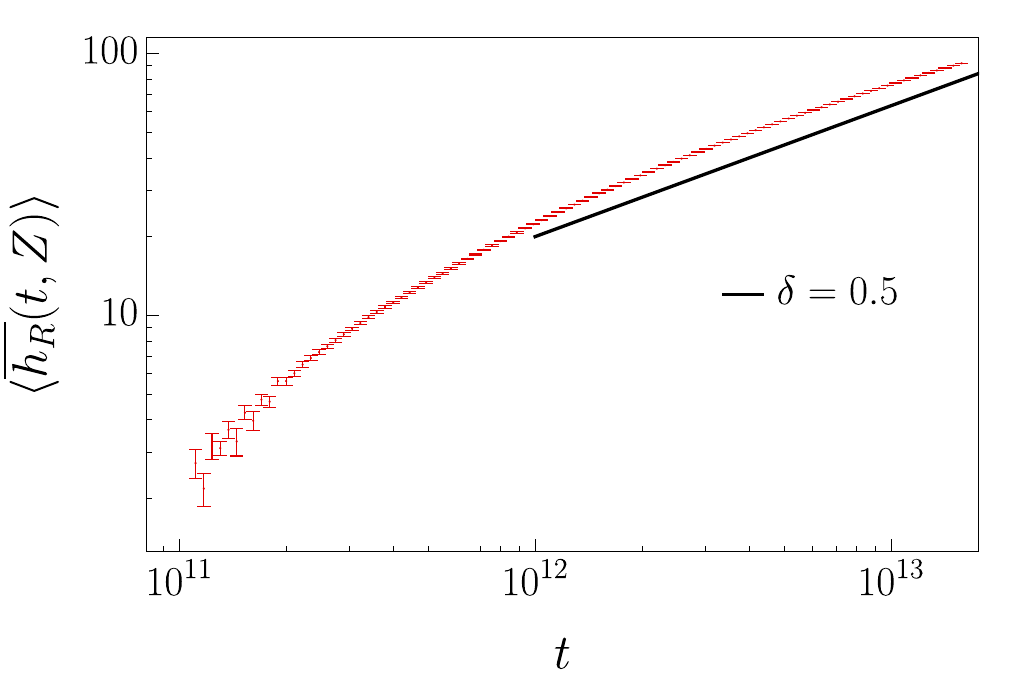}
\caption{(Color online) Average front position $\langle \overline{h_R}(t,Z)\rangle$ as a  function of time for $T=1/3$, $R_R=11$, $Z=1$, and $A=10$. The solid black line corresponds to the reference scaling $\langle \overline{h_R}(t,Z) \rangle \sim t^{1/2}$.}
\label{fig:ht_anomalo}
\end{figure}

Figure \ref{fig:ht_anomalo} depicts the average front position for the most realistic condition studied (smallest temperature and highest Hamaker constant). While it does not reach a steady state of growth, it is in close proximity to the expected diffusive growth, $\langle \overline{h_R}(t,Z) \rangle \sim t^{1/2}$.

\onecolumngrid

\begin{table}[t]
\begin{ruledtabular}
\begin{tabular}{|c|c|c|c|c|c|c|c|c|c|c|}
\multirow{2}{*}{\diagbox[width=2.5em]{$A$}{$T$}} & \multicolumn{2}{c|}{3}  & \multicolumn{2}{c|}{1}  & \multicolumn{2}{c|}{3/4}  & \multicolumn{2}{c|}{1/2} & \multicolumn{2}{c|}{1/3} \\  \cline{2-11}

 & $\delta$ & $2\beta$& $\delta$ & $2\beta$& $\delta$ & $2\beta$& $\delta$ & $2\beta$& $\delta$ & $2\beta$ \\ \hline
 \hline
         10  & 0.285(4) & 0.335(6) & 0.287(3) & 0.315(6) &  0.278(3)& 0.273(7) & 0.253(3) & 0.142(9) & 0.17(3) & 0.13(2) \\ \hline
        3 & 0.315(3) & 0.358(7) & 0.306(3) & 0.360(7) & 0.291(3) & 0.295(8) & 0.247(4) & 0.16(2) & 0.168(5) & 0.14(2) \\ \hline
        1 & 0.312(4) & 0.38(1) & 0.358(3) &  0.459(9) & 0.382(4) & 0.41(9) & 0.367(1) & 0.26(3) & 0.3496(8) & 0.14(2)\\ \hline
        1/3 & 0.340(3) & 0.392(7) & 0.372(2) & 0.44(1) & 0.387(2) & 0.24(4) & 0.353(1) & 0.20(3) & 0.3450(8) & 0.17(3)\\ \hline
        0.1 & 0.341(3) & 0.399(7) & 0.373(2) & 0.44(1) & 0.385(2) & 0.27(2) & 0.3542(8) & 0.24(4) & 0.350(1) & 0.14(2)\\ 
\end{tabular}
\end{ruledtabular}
\caption{Values of the exponents $\delta$ and $2\beta$, for the supernatant layer, for all the conditions studied.}
\label{tab:delta_supernatant}
\end{table}
\begin{table}[t]
\begin{ruledtabular}
\begin{tabular}{|c|c|c|c|c|c|c|c|c|c|c|}
\multirow{2}{*}{\diagbox[width=2.5em]{$A$}{$T$}}  & \multicolumn{2}{c|}{3}  & \multicolumn{2}{c|}{1}  & \multicolumn{2}{c|}{3/4}  & \multicolumn{2}{c|}{1/2} & \multicolumn{2}{c|}{1/3} \\  \cline{2-11}

 & $1/z$ & $2\alpha$&$1/z$ & $2\alpha$& $1/z$ & $2\alpha$& $1/z$ & $2\alpha$& $1/z$ & $2\alpha$ \\ \hline
 \hline
         10 & 0.37(2)& 0.80(7)& 0.29(2)& 0.77(7) & 0.28(2)& 0.56(6)& 0.17(5)& 0.4(1)& 0.37(3)& 0.20(5)\\ \hline
        3 & 0.47(2)& 0.76(3)& 0.21(2)& 1.0(1) & 0.13(2)& 0.94(2)& 0.38(1)& 0.43(2)& 0.38(2)& 0.21(4)\\ \hline
        1 & 0.39(1)& 0.91(3)& 0.35(2)& 1.22(6) & 0.24(2)& 1.5(1)& 0.20(2)& 1.0(2)& \bf{ 0.28(1)}& \bf{ 0.48(6)}\\ \hline
        1/3 & 0.39(1)& 0.89(4)& 0.28(1)& 1.47(7) & 0.18(2)& 1.5(2)& \bf{ 0.26(2)}& \bf{ 0.81(6)}& \bf{ 0.28(1)}& \bf{ 0.52(6)}\\ \hline
        0.1 & 0.39(1)& 0.95(3)& 0.31(2)& 1.24(6) & 0.20(2)& 1.3(2)& \bf{ 0.25(2)} & \bf{ 0.83(6)}& \bf{ 0.28(1)}&\bf{ 0.42(7)}\\ 
\end{tabular}
\end{ruledtabular}
\caption{Values of the exponents $1/z$ and $2\alpha$, for the supernatant layer, for all the conditions studied. In bold those calculated approximating the peak as a parabola.}
\label{tab:z_supernatant}
\end{table}
\begin{table}[b]
\begin{ruledtabular}
\begin{tabular}{|c|c|c|c|c|c|}
\diagbox[width=2.5em]{$A$}{$T$} & 3 & 1 & 3/4 & 1/2 & 1/3\\ \hline
 \hline
        10 & 0.70(7)& 0.73(7)& 0.52(6)& * & $-0.41(6)$ \\ \hline
        3 & 0.58(3)& 1.0(1)& 0.9(2)& * & $-0.41(4)$ \\ \hline
        1 & 0.70(3)& 1.12(6)& 1.3(1)& 0.6(2)& $-0.66(6)$ \\ \hline
        1/3 & 0.68(4)& 1.37(7)& 1.3(2)& 0.30(7)& $-0.60(6)$ \\  \hline
        0.1 & 0.74(3)& 1.13(6)& 1.1(2)& 0.31(6)& $-0.72(6)$ \\ 
\end{tabular}
\end{ruledtabular}
\caption{Value of the exponent $2\alpha'$ for the supernatant layer, for all the conditions studied. We denote with an asterisk two conditions in which the the collapse of the height-difference correlation function was so noisy that was impossible to compute an exponent.}
\label{tab:dap_supernatant}
\end{table}

\newpage
\twocolumngrid
\bibliography{ThinFilm}


\end{document}